\documentclass[journal,onecolumn,12pt,draftclsnofoot]{IEEEtran}
\usepackage[T1]{fontenc}
\usepackage{amsmath,amssymb,amsthm,mathtools,booktabs,array,graphicx}
\usepackage{tikz}
\usetikzlibrary{arrows.meta,positioning,calc}
\usepackage{xcolor,enumitem,cite,needspace}
\usepackage[unicode,hidelinks]{hyperref}
\definecolor{ink}{HTML}{163A50}
\definecolor{blue}{HTML}{246C9D}
\definecolor{orange}{HTML}{A96827}
\definecolor{green}{HTML}{327653}
\definecolor{pale}{HTML}{F0F5F7}
\newcommand{\F}{\mathbb F}
\newcommand{\Acal}{\mathcal A}
\newcommand{\In}{\operatorname{In}}
\newcommand{\Out}{\operatorname{Out}}
\newcommand{\supp}{\operatorname{supp}}
\newcommand{\Span}{\operatorname{span}}
\newcommand{\im}{\operatorname{im}}
\newcommand{\divg}{\operatorname{div}}
\newcommand{\Unif}{\operatorname{Unif}}

\newcolumntype{L}[1]{>{\raggedright\arraybackslash}p{#1}}
\setlist{topsep=4pt,itemsep=2pt,parsep=0pt,leftmargin=2em}
\theoremstyle{definition}
\newtheorem{theorem}{Theorem}
\newtheorem{lemma}[theorem]{Lemma}

\newtheorem{corollary}[theorem]{Corollary}
\newtheorem{definition}{Definition}
\newtheorem*{connectivitylemma}{Connectivity lemma}
\hypersetup{pdftitle={Key Relay and Secure Network Coding: One-Shot Equivalence and Separation on Undirected Networks},pdfauthor={Yuji Ando, Edmund Chen, Go Kato, Toyohiro Tsurumaru, Derek Zhang, and Donald Zvada}}
\title{Key Relay and Secure Network Coding:\\One-Shot Equivalence and Separation\\on Undirected Networks}
\author{Yuji~Ando, Edmund~Chen, Go~Kato,\\
Toyohiro~Tsurumaru, Derek~Zhang, and Donald~Zvada%
\thanks{Yuji Ando is with the Mathematical Science Center for Co-creative Society, Tohoku University, 468-1 Aoba, Aramaki, Aoba-ku, Sendai, Miyagi 980-0845, Japan.}%
\thanks{Edmund Chen is with the Courant Institute of Mathematical Sciences, New York University, USA.
}%
\thanks{Go Kato is with the Quantum ICT Laboratory, NICT, 4-2-1 Nukui-kitamachi, Koganei-shi, Tokyo 184-8795, Japan.}%
\thanks{Toyohiro Tsurumaru is with Mitsubishi Electric Corporation, Information Technology R\&D Center, 5-1-1 Ofuna, Kamakura-shi, Kanagawa 247-8501, Japan.}%
\thanks{Derek Zhang is with the Department of Mathematics, University of Maryland, College Park, Maryland 20742, USA
.}%
\thanks{Donald Zvada is with the O'Reilly Institute, Trinity College Dublin, Dublin 2, D02 N8P5, Ireland.}%
}
\IEEEaftertitletext{\vspace{-3\baselineskip}}
\begin{document}
\maketitle
\begin{abstract}
\setlength{\baselineskip}{0.85\baselineskip}
Motivated by quantum key distribution networks, we compare key relay protocols with secure network coding on the same undirected graph. In this comparison, we fix the finite field, terminal set, and wiretap family. Each edge supplies one independent uniform symbol to its endpoints for key relay, or carries one symbol once for network coding. The network code may choose its source terminal, edge orientations, and causal transmission order before execution. In this setting, the standard one-time-pad simulation readily converts every secure network code to key relay. We investigate the reverse conversion to linear network coding. For two terminals and a one-symbol key, arbitrary key relay is equivalent to scalar linear acyclic network coding over the prescribed field. The proof uses connectivity and flow conservation, related to the binary $st$-flow protocol. However, this equivalence fails in general when several terminals seek one common key. A doubled triangle separates key relay from acyclic coding but admits causal coding on a cyclic orientation. On the four-terminal complete graph, a linear key relay protocol over the ternary field withstands the three perfect-matching wiretap sets, whereas no linear causal network code distributes a source symbol of the same size, even with private randomness at every vertex and its use in decoding. We also establish a linear key relay normal form that removes interaction and private randomness and bounds public communication. All results concern one-shot feasibility with exact agreement and perfect secrecy.
\end{abstract}
\begin{IEEEkeywords}
Key relay, pairwise independent networks, secret-key agreement, secure network coding.
\end{IEEEkeywords}
\newpage
\section{Introduction}
Quantum key distribution (QKD) establishes a shared secret key between two endpoints connected by a quantum channel and an authenticated classical channel. A QKD network connects such links to serve multiple locations. This motivates a basic problem in secret-key agreement: how can keys shared between neighboring nodes be used to establish a common key at distant terminals? A \emph{key relay protocol} (KRP) combines these local keys with public discussion. The term \emph{key relay} is used in QKD network literature and terminology~\cite{Elliott2004,ITUY3800}. We separate the generation of link keys from their subsequent use and idealize the initial resource as independent random symbols, each shared identically by the endpoints of one edge. The edge-eavesdropping model below represents disclosure of selected link keys at this classical protocol layer.

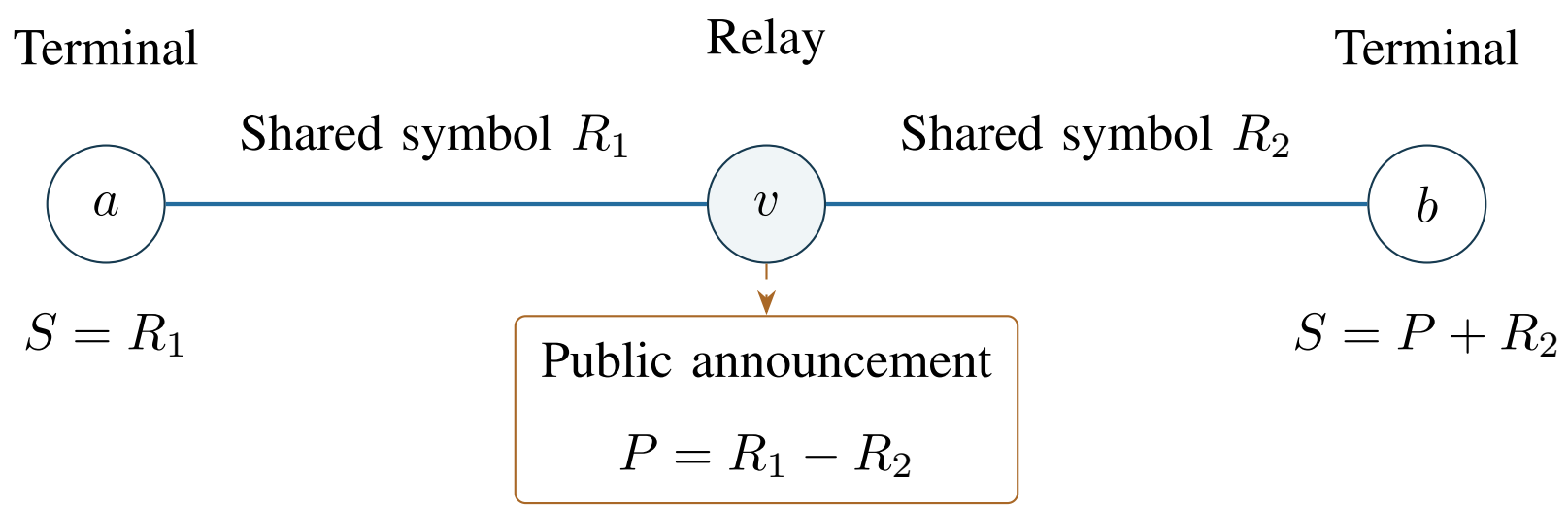
\begin{figure}[t]
\centering
\begin{tikzpicture}[>=Stealth,vertex/.style={circle,draw=ink,fill=white,minimum size=8mm,inner sep=0pt},edge/.style={draw=blue,line width=.8pt},every node/.style={font=\small}]
\node[vertex] (a) at (0,0) {$a$};
\node[vertex,fill=pale] (v) at (4.5,0) {$v$};
\node[vertex] (b) at (9,0) {$b$};
\draw[edge] (a)--node[above=4pt] {Shared symbol $R_1$}(v);
\draw[edge] (v)--node[above=4pt] {Shared symbol $R_2$}(b);
\node[above=8mm] at (a) {Terminal};
\node[above=8mm] at (v) {Relay};
\node[above=8mm] at (b) {Terminal};
\node[below=6mm] at (a) {$S=R_1$};
\node[below=6mm] at (b) {$S=P+R_2$};
\node[draw=orange,rounded corners=2pt,align=center,inner sep=5pt] (p) at (4.5,-1.4) {Public announcement\\$P=R_1-R_2$};
\draw[->,draw=orange,dashed] (v)--(p);
\end{tikzpicture}
\caption{Key relay on a path. Each edge supplies randomness to both endpoints; it does not specify a transmission direction. The public announcement alone reveals no information about the common key.}
\label{fig:krp-path}
\end{figure}

For example, consider the path $a-v-b$ in Fig.~\ref{fig:krp-path}. The edges supply independent uniform symbols $R_1,R_2\in F$. The relay announces $P=R_1-R_2$, after which $a$ and $b$ share $S=R_1=P+R_2$. The announcement is independent of the key. This example assumes that no edge randomness is exposed; our general model also requires secrecy when an adversary learns the random symbols on a permitted set of edges.

Secure network coding (SNC) distributes a secret message held by a designated source through values computed and transmitted along network edges~\cite{CaiYeung2011}. Its extensive theory makes the following question important for key relay: whenever a KRP establishes a common key, can SNC securely distribute a source message of the same size to the same terminals using the corresponding network resources? If so, studying key relay would largely reduce to studying the corresponding SNC problem. Shared link keys do not, however, prescribe transmission directions. To test this possible reduction, SNC should be allowed to choose its source terminal and edge directions, rather than being confined to one prescribed directed graph.

We make this comparison on a fixed undirected physical graph, with a fixed finite field, terminal set, and family of wiretap sets. In KRP, each edge supplies one independent uniform field symbol to its endpoints, and finite public discussion is allowed. In SNC, each edge carries at most one field symbol, and no free public discussion is added. The SNC designer may choose any terminal as the source, any orientation of the used edges, and any causal transmission order; these choices are fixed before execution. One scheme must work for every allowed wiretap set. The eavesdropper observes the full public transcript and the exposed edge symbols in KRP, or the tapped transmissions in SNC. This one-shot comparison isolates the use of a fixed supply of link-key symbols and the corresponding one-use communication resources.

The published comparison of Kato, Fujiwara, and Tsurumaru~\cite{Kato2023} provides our starting point. They show that one-time-pad simulation converts SNC to KRP, prove equivalence when free public channels are added to SNC, and exhibit a strict advantage of KRP over SNC for several independent keys requested by different terminal pairs. Their model already allows nonlinear causal operations. They conjecture that the gap disappears for a single sender--receiver pair~\cite[Sec.~VI]{Kato2023}. We revisit this question for a one-symbol key in the finite, fixed-schedule model defined below, and establish equivalence among arbitrary KRP, linear KRP, and scalar linear acyclic SNC over the same prescribed field. Our central separation concerns several terminals requesting \emph{one common key}: a four-terminal example excludes every linear causal SNC in our model, for every choice of source, orientation, and transmission order, even when all vertices have private randomness and receivers may use it in decoding.

Considering both multiple key terminals and nonbinary alphabets is consequential for this comparison. The two-terminal equivalence with linear SNC does not extend to arbitrary terminal sets, and allowing the ternary field provides the explicit four-terminal KRP used in our separation. These choices let us examine where the equivalence with linear SNC ceases to hold while keeping the physical graph, edge alphabet, and secrecy requirement the same on both sides.

\subsection{Connections to Secret-Key Agreement and Network Coding}
Secret-key agreement from correlated observations and public discussion provides a broader context for KRP~\cite{CsiszarNarayan2004}. In the basic pairwise independent network (PIN) key-agreement problem~\cite{PIN2010}, the eavesdropper observes the public discussion but has no additional access to the pairwise source observations. In our notation, this secrecy setting corresponds to $\Acal=\{\varnothing\}$: no edge keys are exposed. When the two endpoints of each edge observe the same independent uniform symbol, the initial resources also coincide with those of KRP. Thus the noiseless uniform-edge specialization of PIN and the no-edge-leakage specialization of KRP overlap. Perfect key generation from noiseless PIN sources was studied by Nitinawarat and Narayan~\cite{PerfectPIN2010}. Our general KRP model additionally requires one protocol to remain perfectly secret for every allowed exposed-edge set, given both the leaked edge keys and the public discussion. We retain the name KRP for this QKD-motivated protocol and compare one-shot feasibility with exact agreement and perfect secrecy.

For the case without edge leakage, Chan proved equality between secret-key capacity for a noiseless PIN source and multicast capacity on the corresponding undirected network~\cite{ChanHidden2011,ChanMatroidal2012}. In this equivalence, the eavesdropper observes only public discussion in the source model, while the network-coding model has no edge wiretapping. This capacity result does not itself give an equivalence preserving our prescribed wiretap family and one-use, fixed-alphabet edge resources.

Mizuki, Nakayama, and Sone introduced a \emph{one-time network} model for two-terminal binary key agreement on partially leaked key-sharing graphs~\cite{Mizuki2011}. They specify a probability distribution on the exposed-edge sets and maximize the conditional entropy of the shared bit given the eavesdropper's observations. If the distribution assigns positive probability to every set in a prescribed family and zero probability outside it, zero average information leakage is equivalent to perfect secrecy for every set in that family. Thus, for two terminals, one-bit edge keys, and a one-bit common key, our KRP secrecy requirement corresponds to the zero-leakage case of their formulation.

Their binary $st$-flow protocol also provides a direct antecedent for our two-terminal proof. Lemmas~8 and~10 of~\cite{Mizuki2011} give perfect secrecy when the exposed edges do not separate the terminals and impossibility for any protocol when they do. The same binary construction can be interpreted as scalar linear acyclic SNC (model D0, defined in Section~\ref{sec:snc}) by replacing its one-time-pad public transmissions with transmissions along the corresponding edges. Section~\ref{sec:two} explains this correspondence and proves the KRP--SNC equivalence over any prescribed finite field. The comparison with SNC and the case of three or more key terminals are not addressed in~\cite{Mizuki2011}.

In a related but distinct comparison, Langberg and Effros~\cite{Keycast2025} show that key-cast can outperform communication requiring reconstruction of the original source information in directed noiseless networks. Their comparison takes place within that communication model; it does not compare SNC with KRP based on preshared edge keys and public discussion.

On the SNC side, the literature includes source-generated randomness~\cite{CaiYeung2011,CaiChan2011}, as well as restricted wiretap families and randomness at intermediate vertices~\cite{CuiHoKliewer2013}. Our four-terminal example uses the three perfect matchings as maximal wiretap sets, rather than every pair of edges. We distinguish acyclic coding, causal coding on a possibly cyclic orientation, and decoding that uses receiver randomness; Section~\ref{sec:snc} gives their precise definitions. We focus on linear SNC when analyzing the general reverse conversion. Several results extend beyond that restriction: the SNC-to-KRP simulation permits nonlinear codes, the two-terminal equivalence permits arbitrary KRP, and the doubled-triangle converse excludes nonlinear D0 codes.

For finite linear sources, Chan \emph{et al.}~\cite{Chan2019} study one-shot perfect key agreement with linear discussion, explaining how to remove linear interaction and why private randomization is unnecessary in their setting. Our linear normal form uses these principles while preserving secrecy for the entire wiretap family, allowing helper nodes, and bounding each node's announcement by $(r-1)\ell$ symbols for $r$ terminals and an $\ell$-symbol key. We use the normal form as a structural tool for the resource comparison.

\subsection{Contributions and Organization}
We distinguish three SNC models: a directed acyclic model D0, a causal edge-ordered model D1, and a model D2 that additionally permits each receiver to use its own local randomness in decoding. All permit randomness at intermediate nodes. Our results are as follows.
\begin{enumerate}
\item Every admissible D0, D1, or D2 code, including a nonlinear code, yields an admissible KRP by chronological one-time-pad simulation (Theorem~\ref{thm:otp}). This gives the common direction of comparison using the established simulation construction.
\item A linear KRP with finite public discussion has a noninteractive realization without private randomness and with at most $(r-1)\ell$ public symbols per node (Theorem~\ref{thm:normal}).
\item For two terminals and a one-symbol key, arbitrary KRP, linear KRP, and scalar linear D0 SNC are equivalent over the same prescribed field $F$ (Theorem~\ref{thm:two}). The proof uses connectivity after removal of each allowed wiretap set and gives an explicit causal realization without enlarging any edge alphabet. Other vertices may act as helpers.
\item A doubled triangle separates KRP from D0 even for nonlinear codes, but admits D1 SNC (Theorem~\ref{thm:triangle}). Thus a restriction to acyclic orientations can itself create a gap.
\item On the complete graph with four terminals and the three perfect matchings as maximal wiretap sets, a linear KRP over $\F_3$ shares one symbol, whereas no linear D2 SNC distributes a source symbol of the same size. This holds for every source, orientation, and causal edge order, even when all vertices have private randomness and receivers may use it in decoding (Theorem~\ref{thm:k4}).
\end{enumerate}
Sections~\ref{sec:krp} and~\ref{sec:snc} define the resources and simulation. Section~\ref{sec:linear} gives the linear normal form. Section~\ref{sec:two} proves the two-terminal equivalence, and Sections~\ref{sec:triangle} and~\ref{sec:k4} give the two separations. The four-terminal proof identifies a conflict between local linear relations and causal generation of the edge values. Appendices~\ref{app:normal} and~\ref{app:local} supply the normal-form and local-relation proofs.

\section{Key Relay Protocols}
\label{sec:krp}
We compare KRP and SNC on the same physical graph, with the same edge alphabet, terminal set, and family of wiretap sets. We first specify their common eavesdropping model and then define KRP.

\subsection{Graph and Eavesdropping Model}
All entropies and mutual informations are measured in bits, and $\log$ denotes the base-two logarithm.

Fix a finite loopless undirected multigraph $G=(V,E)$, a finite field $F$, and a set $U\subseteq V$ of at least two terminals. Parallel edges are distinct physical resources. Let $E(v)$ denote the set of edges incident to $v$. These physical edge identities are retained when directions and a transmission order are chosen for SNC.

\paragraph{Notation and dimensions}
An indexed collection such as $R_B=(R_e)_{e\in B}$ or $Y_B=(Y_e)_{e\in B}$ is arranged in a fixed order and regarded as a column vector. Collections of vectors, including $K_V$ and the public transcript $P$, are stacked vertically in the same way. We write $F^d$ for column vectors with $d$ entries, allowing $d=0$ for an empty vector. Uppercase italic letters denote random variables, which may be scalar or vector valued; their dimensions are specified when they are introduced. In the linear-algebra arguments, deterministic matrices are written in sans-serif capitals, such as $\mathsf A$, and coefficient row vectors in bold lowercase, such as $\mathbf g_e$. In particular, a coefficient row vector and a random column vector are different types of objects. Table~\ref{tab:notation} collects the principal notation used in the two protocol models.

\begin{table}[t]
\caption{Principal notation. Each vector is a column vector unless explicitly described as a coefficient row vector.}
\label{tab:notation}
\centering\small
\renewcommand{\arraystretch}{1.15}
\begin{tabular}{L{37mm}L{113mm}}
\toprule
Symbol & Meaning and type \\
\midrule
$G=(V,E)$, $U\subseteq V$ & Physical graph and set of key terminals; other vertices may act as helpers. \\
$F$, $\ell$ & Finite field and number of symbols in the shared key or source message. \\
$\Acal\subseteq 2^E$ & Family of permitted wiretap sets; one scheme must be secure for every member. \\
$R_e$, $R_B$ & Shared edge symbol in $F$, and the column of edge symbols indexed by $B\subseteq E$. \\
$K_v$ & Vertex $v$'s private random column vector; its length may depend on $v$. \\
$P_t$, $P$ & A public-message column vector, and the column stacking all public messages. \\
$S$, $M$ & KRP common key and SNC source message, respectively, in $F^\ell$. \\
$Y_e$, $Y_B$ & SNC transmitted symbol in $F$, and the column of transmissions indexed by $B$. \\
$\mathsf A$, $\mathbf g$ & A deterministic matrix and a coefficient row vector, respectively; sizes are stated locally. \\
\bottomrule
\end{tabular}
\end{table}

\Needspace{8\baselineskip}
\begin{definition}[Wiretap sets]
\label{def:wiretap}
A subset $A\subseteq E$ is a \emph{wiretap set}, and a family $\Acal\subseteq 2^E$ specifies the sets available to the eavesdropper. The eavesdropper chooses one $A\in\Acal$ and observes all information associated with the edges in $A$. A scheme may depend on $\Acal$, but not on the particular choice of $A$, and must be secure for every $A\in\Acal$. We assume $\varnothing\in\Acal$. Eavesdropping is passive, and every vertex follows the prescribed protocol.
\end{definition}

This definition applies to KRP and all the SNC models below. The information associated with an edge differs between the two settings: the eavesdropper observes the shared edge variables $R_A$ and the entire public transcript $P$ in KRP, and the transmitted edge values $Y_A$ in SNC. Thus, its observations are $(P,R_A)$ and $Y_A$, respectively. Replacing $\Acal$ by its downward closure does not change the secrecy requirement. In particular, resistance to observation of any $k$ edges is described by the threshold family
\[
 \Acal_k=\{A\subseteq E:|A|\le k\}.
\]

\subsection{Operation and Admissibility}
\Needspace{8\baselineskip}
\begin{definition}[KRP]
\label{def:krp}
For a key length $\ell\ge1$, a \emph{KRP} on $(G,U,F,\Acal)$ consists of the following resources and operations.
\begin{enumerate}[label=(\roman*),leftmargin=2em]
\item \emph{Initial information:} Each edge $e$ provides a uniform random variable $R_e\in F$ to its two endpoints. Each vertex $v$ may also have a finite vector $K_v$ of private random variables. All components of $R_E$ and $K_V$ are mutually independent and uniform over $F$.
\item \emph{Public discussion:} The protocol has finitely many public messages. The speaker sequence $v_1,\ldots,v_T$ and all message lengths are fixed independently of the data; a vertex may appear repeatedly. At time $t$, vertex $v_t$ announces a finite vector
\begin{equation}
 P_t=\psi_t\bigl(R_{E(v_t)},K_{v_t},P_{<t}\bigr),
 \label{eq:krp_public}
\end{equation}
where $P_{<t}=(P_1,\ldots,P_{t-1})$. The full transcript $P=(P_1,\ldots,P_T)$ is available to every vertex and to the eavesdropper. There is no common upper bound on the total amount of public discussion. We allow $T=0$ and messages of length zero.
\item \emph{Outputs:} Each terminal $u\in U$ outputs
\begin{equation}
 S_u=\rho_u\bigl(R_{E(u)},K_u,P\bigr)\in F^\ell.
 \label{eq:krp_decode}
\end{equation}
\end{enumerate}
\end{definition}

\Needspace{8\baselineskip}
\begin{definition}[Admissible KRP]
\label{def:admissible-krp}
A KRP is \emph{admissible} if it satisfies perfect agreement, uniformity, and perfect secrecy, respectively:
\begin{align}
 &S_u=S\quad(u\in U)\quad\text{for all initial values},
 \label{eq:krp_correct}\\
 &S\sim\Unif(F^\ell),\nonumber\\
 &I(S;P,R_A)=0\quad(A\in\Acal).
 \label{eq:krp_security}
\end{align}
\end{definition}

To express this model in source terminology, write $X_v=R_{E(v)}$ for vertex $v$'s observation and $Z_A=R_A$ for the eavesdropper's side information. The observations $(X_v:v\in V)$ form a noiseless PIN source with independent uniform edge variables~\cite{PIN2010,PerfectPIN2010}. The set $U$ consists of the key terminals, the remaining vertices act as helpers, and $K_v$ supplies independent private randomization. The same discussion and decoding rules must produce a uniform common key satisfying $I(S;P,Z_A)=0$ for every $A\in\Acal$. The eavesdropper observes $Z_A$ for its chosen set $A$.

Uniformity excludes a constant output and fixes the key size at $\ell$ field symbols. The protocol in Fig.~\ref{fig:krp-path} is admissible for $U=\{a,b\}$, $\ell=1$, and $\Acal=\{\varnothing\}$. From Section~\ref{sec:two} onward, we take $\ell=1$ unless stated otherwise.

\section{Secure Network Coding and Its Simulation by KRP}
\label{sec:snc}
SNC also distributes a secret over a network. Here, each edge is a communication channel, and a source delivers an initially held message to the other terminals. We define three models, D0, D1, and D2, and show that KRP can simulate each of them, without any linearity assumption.

\subsection{Three SNC Models: D0, D1, and D2}
\label{subsec:snc-models}
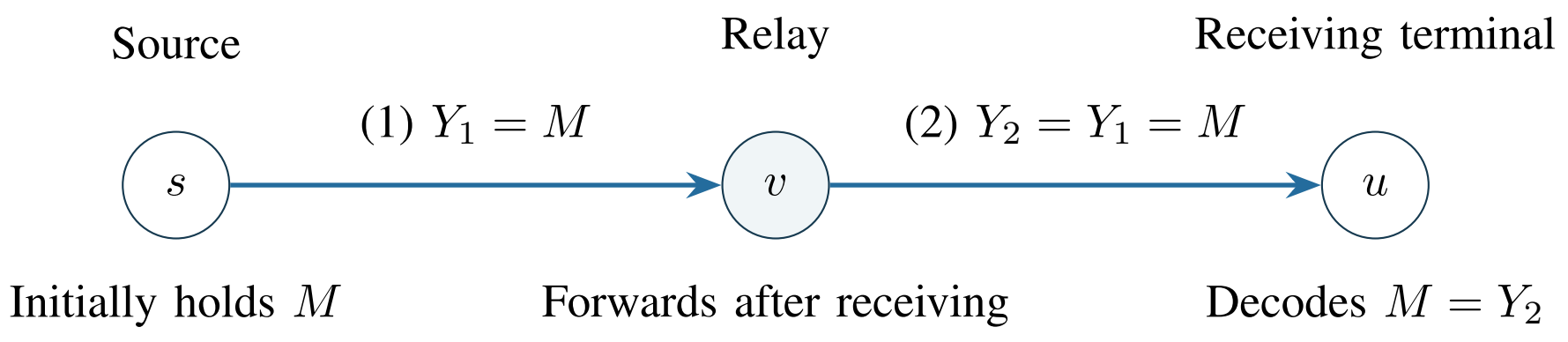
\begin{figure}[t]
\centering
\begin{tikzpicture}[
  >=Stealth,
  vertex/.style={circle,draw=ink,fill=white,minimum size=8mm,inner sep=0pt},
  edge/.style={->,draw=blue,line width=1.0pt},
  every node/.style={font=\small}
]
  \node[vertex] (s) at (0,0) {$s$};
  \node[vertex,fill=pale] (v) at (4.5,0) {$v$};
  \node[vertex] (u) at (9,0) {$u$};
  \draw[edge] (s) -- node[above=4pt] {(1) $Y_1=M$} (v);
  \draw[edge] (v) -- node[above=4pt] {(2) $Y_2=Y_1=M$} (u);
  \node[above=8mm] at (s) {Source};
  \node[above=8mm] at (v) {Relay};
  \node[above=8mm] at (u) {Receiving terminal};
  \node[below=6mm,align=center] at (s) {Initially holds $M$};
  \node[below=6mm,align=center] at (v) {Forwards after receiving};
  \node[below=6mm,align=center] at (u) {Decodes $M=Y_2$};
\end{tikzpicture}
\caption{DAG-based SNC (D0) on a path. Each arrow represents one use of a physical edge in one direction. Edge values are computed by the senders; they are not initially shared random variables.}
\label{fig:snc-path}
\end{figure}

The simplest example is the path $s\to v\to u$ in Fig.~\ref{fig:snc-path}. The source $s$ sends a message $M$, and the relay $v$ forwards it to $u$. As in Fig.~\ref{fig:krp-path}, this example uses $\Acal=\{\varnothing\}$. Observation of either edge reveals $M$, so this path protocol does not protect against any nonempty wiretap set.

In general, fix the same $(G,U,F,\Acal)$ as for KRP and a message length $\ell$. We first specify the common resources and then distinguish the encoding and decoding rules.

\Needspace{8\baselineskip}
\begin{definition}[Common SNC resources]
\label{def:snc-resources}
A source $s\in U$ holds a uniform message $M\in F^\ell$, and the terminals in $U\setminus\{s\}$ are receivers. Each vertex $v$ may have a finite vector $K_v$ of private random variables. All components of $K_V$ are mutually independent, uniform over $F$, and independent of $M$. Each physical edge is used at most once, in a direction fixed independently of the data, to transmit one symbol of $F$. No free public discussion is available. Write $Y_e$ for the value transmitted on edge $e$, and set $Y_e=0$ for every unused edge. Among the used edges, let $\In(v)$ and $\Out(v)$ denote the edges entering and leaving $v$, respectively.
\end{definition}

The source, used edges, and their directions are design choices, not additional prescribed inputs to the undirected instance. For the causal models below, the transmission order is also chosen by the designer. An existence statement permits these choices; an impossibility statement covers all of them. A single choice of code and schedule must satisfy the requirements for every $A\in\Acal$.

Private randomness at intermediate vertices has been considered in prior work. In particular, \cite[Sec.~II]{CuiHoKliewer2013} defines a DAG model in which a vertex encodes using its private randomness after receiving all its inputs. Models in which only the source generates randomness have also been studied extensively \cite{CaiYeung2011}. The placement of randomness and the linearity of encoding or decoding are separate assumptions \cite{CaiHayashi2020}.

\Needspace{8\baselineskip}
\begin{definition}[DAG-based SNC: D0]
\label{def:dag}
Under the common resources above, suppose that the directed graph of used edges is acyclic. Each vertex transmits after receiving all its inputs. For an edge $e:v\to w$, its transmitted value is
\begin{equation}
 Y_e=\begin{cases}
 \phi_e\bigl(Y_{\In(v)},K_v\bigr),&v\ne s,\\
 \phi_e(M,K_s),&v=s.
 \end{cases}
 \label{eq:dag_encoder}
\end{equation}
The source does not use received values. Each receiving terminal decodes from $Y_{\In(u)}$ alone. We call this model \emph{D0}.
\end{definition}

We next remove the acyclicity requirement and permit a transmission whenever its required inputs have arrived. An ordering of channel encodings is used in \cite[Sec.~II]{CaiYeung2007}; the model in \cite{HayashiOwariKatoCai2020} permits encoding from previously received values without assuming an acyclic graph. The SNC model in \cite{Kato2023} also uses each confidential channel once and allows transmission when the required information is available. The following definition makes the causal and resource assumptions needed for our comparison explicit.

\Needspace{8\baselineskip}
\begin{definition}[Causal SNC: D1 and D2]
\label{def:causal}
Under the common resources above, fix an ordering $e_1,\ldots,e_N$ of the used edges independently of the data. Directed cycles are allowed, but each transmission may depend only on previously received values. If $v_i$ sends on $e_i$, let $\In_{<i}(v_i)$ be the edges on which $v_i$ has received before time $i$. Then
\begin{equation}
 Y_{e_i}=\begin{cases}
 \phi_i\bigl(Y_{\In_{<i}(v_i)},K_{v_i}\bigr),&v_i\ne s,\\
 \phi_i\bigl(M,Y_{\In_{<i}(s)},K_s\bigr),&v_i=s.
 \end{cases}
 \label{eq:causal_encoder}
\end{equation}
Thus, the source may also use previously received values. We call the model \emph{D1} if each receiving terminal decodes from its received values alone, and \emph{D2} if it may additionally use its own private randomness.
\end{definition}

\Needspace{8\baselineskip}
\begin{definition}[Admissible SNC]
\label{def:admissible-snc}
An SNC in any of the models D0, D1, or D2 is \emph{admissible} if it satisfies perfect secrecy,
\begin{equation}
 I(M;Y_A)=0\quad(A\in\Acal),
 \label{eq:snc_security}
\end{equation}
and the prescribed perfect decoding condition at every $u\in U\setminus\{s\}$:
\begin{align}
 \mathrm{D0},\mathrm{D1}:&\quad H(M\mid Y_{\In(u)})=0,\nonumber\\
 \mathrm{D2}:&\quad H(M\mid Y_{\In(u)},K_u)=0.
 \label{eq:d1d2}
\end{align}
No linearity is required. An SNC is \emph{linear} if all its encoding and decoding maps are linear over $F$.
\end{definition}

\paragraph{Relations among the models}
A D0 code can be executed in D1 by ordering the outgoing edges according to a topological order of their senders. A D1 code is also a D2 code whose decoders do not use private randomness. Hence
\begin{equation}
 \mathrm{D0}\subseteq\mathrm{D1}\subseteq\mathrm{D2}.
 \label{eq:model-inclusion}
\end{equation}
These names refer to the entire models, not merely their decoding rules. In a DAG, $K_u$ is independent of $(M,Y_{\In(u)})$, so allowing $K_u$ at the decoder does not improve decodability. This argument does not apply in the presence of directed cycles, because the influence of a vertex's private randomness can return in its later received values.

For example, consider two parallel edges used in the order $u\to s:R$, followed by $s\to u:M+R$. Terminal $u$ subtracts its private random symbol $R$ to recover $M$. This decoder is allowed in D2, whereas $M+R$ alone does not determine $M$. Each individual edge reveals no information about $M$. Receiver randomness and communication in the reverse direction also appear in \cite{HuangHoLangbergKliewer2014}; that work uses a capacity model allowing communication over multiple time steps, whereas each physical edge here can be used only once.

\paragraph{Resources preserved in the comparison}
We preserve the physical edges, field $F$, terminal set $U$, wiretap family $\Acal$, and key or message length $\ell$. Splitting an edge or using it repeatedly to increase SNC communication resources is not allowed. KRP has no common upper bound on the number or total length of its public messages, and the same edge variable may appear in several public-message functions. Thus ``one-shot'' refers to one initial symbol per edge and one transmission per edge on the respective sides, not to a single round of public discussion. We compare the feasibility of distributing one common secret of the specified size; we do not require equality of the individual public messages or of the resulting key random variables. Table~\ref{tab:models} summarizes the models. For background on ordinary network coding, see \cite{Yeung2006}.

\begin{table}[t]
\caption{The four models compared on common physical resources. Independent private randomness is allowed in every model. The SNC decoding inputs are those of a receiving terminal. All edge directions and communication schedules are fixed independently of the data.}
\label{tab:models}
\centering\footnotesize
\setlength{\tabcolsep}{3.5pt}
\renewcommand{\arraystretch}{1.2}
\begin{tabular}{L{22mm}L{32mm}L{32mm}L{31mm}L{31mm}}
\toprule
Feature & KRP & D0: DAG-based & D1: causal & D2: causal\\
\midrule
Shared secret & Generated key $S$ & Source message $M$ & Source message $M$ & Source message $M$\\
Resource per edge & Shared random symbol $R_e$ & At most one symbol in one direction & Same as D0 & Same as D0\\
Public discussion & Finitely many messages; repeated speakers allowed & None & None & None\\
Communication order & Fixed speaker sequence & Topological vertex order; all inputs received before encoding & Fixed edge order; only earlier received values used & Same as D1\\
Source uses received values & No designated source & No & Allowed & Allowed\\
Decoding inputs & $R_{E(u)},K_u,P$ & $Y_{\In(u)}$ & $Y_{\In(u)}$ & $Y_{\In(u)},K_u$\\
Eavesdropper observes & $(P,R_A)$ & $Y_A$ & $Y_A$ & $Y_A$\\
\bottomrule
\end{tabular}
\end{table}

\subsection{One-Time-Pad Simulation of SNC by KRP}
\label{subsec:otp}
Replacing a channel transmission by a public one-time-pad ciphertext using shared edge randomness is an established simulation principle~\cite{Kato2023,LaiHo2015}. In particular, Lai and Ho~\cite{LaiHo2015} combine network coding with hop-by-hop one-time-pad encryption using local keys to construct group-key protocols secure against observation of public discussion. We state the simulation for the three causal models above, including decoding with receiver randomness, and explicitly preserve perfect secrecy for the entire prescribed wiretap family.
\begin{theorem}[One-time-pad simulation]
\label{thm:otp}
If an admissible SNC exists in D0, D1, or D2, then an admissible KRP exists on the same resources. Encoding and decoding may be nonlinear. If the SNC is linear, the resulting KRP is also linear.
\end{theorem}
\begin{proof}
By \eqref{eq:model-inclusion}, it suffices to treat D2. In the KRP, the source $s$ generates $M$ as private randomness and simulates the SNC in its prescribed edge order. Instead of transmitting $Y_e$ on $e:v\to w$, vertex $v$ publicly announces the one-time-pad ciphertext
\[
 D_e=Y_e+R_e.
\]
Vertex $w$ subtracts the shared symbol $R_e$ to recover $Y_e$. Each sender has the received values and private randomness required for the original encoding operation. Repeated announcements by the same vertex are permitted by Definition~\ref{def:krp}. Each receiving terminal applies the original D2 decoder to its recovered inputs and private randomness. Thus all terminals, including the source, obtain $S=M$.

For uniform notation, set $Y_e=0$ on unused edges and let one endpoint of each such edge announce $D_e=R_e$ after the simulation. The complete plaintext vector $Y$ depends only on the original SNC variables $(M,K_V)$ and is independent of $R_E$. Consequently, $D=Y+R_E$ is uniform and independent of $(M,Y)$. Moreover, $(D,R_A)$ and $(D,Y_A)$ determine each other, so
\begin{align*}
 I(S;D,R_A)&=I(M;D,Y_A)\\
           &=I(M;Y_A)=0.
\end{align*}
Agreement, uniformity, and perfect secrecy are therefore preserved. If the SNC is linear, all announcements and decoders in the construction are linear.
\end{proof}

\paragraph{The path example}
Applying Theorem~\ref{thm:otp} to Fig.~\ref{fig:snc-path}, vertices $s$ and $v$ announce, respectively,
\[
 D_1=M+R_1,\qquad D_2=M+R_2,
\]
and $u$ recovers $M=D_2-R_2$. The transcript is equivalently represented by $(D_1,P)$, where
\begin{equation}
 P=D_1-D_2=R_1-R_2.
 \label{eq:path-otp}
\end{equation}
Replace the common key $M$ by $S'=D_1-M=R_1$. Conditional on the transcript, this is an invertible transformation of the key and preserves uniformity and secrecy. Now $s$ can output $R_1$, $u$ can output $P+R_2$, and $v$ can compute $P$ from its incident edge variables alone. Omitting the unnecessary announcement $D_1$ and auxiliary randomness $M$ gives exactly the KRP in Fig.~\ref{fig:krp-path}, with $s=a$ and $u=b$. This example illustrates how shared edge randomness and public discussion replace SNC transmissions.

\section{A Normal Form for Linear KRP}
\label{sec:linear}
The simulation in Section~\ref{sec:snc} does not require linearity. We now investigate the converse: when can SNC realize key sharing achievable by KRP? We first use the algebraic structure of linear protocols to study both the possibility and the limitations of this conversion. Linear SNC is also a central subject of the existing literature~\cite{CaiChan2011}. The reduction below builds on the linear-discussion principles of~\cite{Chan2019}; its particular local form preserves the full wiretap family. We state explicitly whenever a conclusion extends beyond linear schemes.

\Needspace{8\baselineskip}
\begin{definition}[Linear KRP]
\label{def:linear-krp}
A KRP is \emph{linear} if its announcements and key are vectors over $F$, and all announcement maps in \eqref{eq:krp_public} and decoding maps in \eqref{eq:krp_decode} are linear over $F$. The speaker sequence and the coefficients of all linear maps are fixed in advance.
\end{definition}

\begin{theorem}[Eliminating private randomness and interaction]
\label{thm:normal}
Let $r=|U|\ge2$. Suppose an admissible linear KRP produces a uniform key in $F^\ell$. For any reference terminal $u_0\in U$, there is an admissible linear KRP on the same resources of the form
\begin{align}
 P'_v&=\mathsf{B}'_vR_{E(v)},\nonumber\\
 S'&=\mathsf{A}_{u_0}R_{E(u_0)},\nonumber\\
 S'&=\mathsf{A}_uR_{E(u)}+\sum_{v\in V}P'_{v,u}
       \quad(u\in U\setminus\{u_0\}),
 \label{eq:normal}
\end{align}
where each $P'_{v,u}\in F^\ell$ is a column vector and $P'_v\in F^{(r-1)\ell}$ is obtained by stacking these blocks over $u\in U\setminus\{u_0\}$ in a fixed order. The deterministic coefficient matrices have dimensions
\[
 \mathsf{A}_u\in F^{\ell\times|E(u)|},\qquad
 \mathsf{B}'_v\in F^{(r-1)\ell\times|E(v)|}.
\]
Thus $R_{E(v)}$ is the column vector of the $|E(v)|$ edge symbols known to $v$, and every product in \eqref{eq:normal} is an ordinary matrix--vector product. Private randomness and interaction are unnecessary: every vertex need only announce once, using at most $(r-1)\ell$ field symbols.
\end{theorem}

For each receiving terminal $u\ne u_0$, the block $P'_{v,u}$ is the contribution from vertex $v$ to that terminal's decoding formula. Vertex $v$ publishes all these blocks as one vector $P'_v$; they need not be transmitted separately. The reference terminal computes the key from its incident edge variables alone, while every other terminal adds the relevant public blocks to its own local term.

The proof uses three linear operations. First, subtracting the contribution of earlier announcements removes interaction, and all resulting announcements from one vertex are grouped together. Second, comparing the decoding identities of two terminals expresses the private-randomness term in a decoder using incident edge variables and public information. Third, subtracting the public term in the reference terminal's decoder translates the common key. The key remains uniform under every fixed eavesdropper observation, so this public translation preserves secrecy. Retaining only the public blocks required by the decoders yields \eqref{eq:normal}. Appendix~\ref{app:normal} gives the matrix calculations and their dimensions.

\begin{corollary}[Two terminals]
\label{cor:two_normal}
For $U=\{s,t\}$ and $\ell=1$, an admissible linear KRP can be replaced by an admissible linear KRP with no private randomness, no announcements by $s$ or $t$, and at most one announced symbol $p_v(R_{E(v)})$ at each relay. Its key has the form
\begin{equation}
 S=\mathsf{A}_sR_{E(s)}
   =\mathsf{A}_tR_{E(t)}+\sum_{v\ne s,t}p_v(R_{E(v)}).
 \label{eq:two_normal}
\end{equation}
Here $\mathsf{A}_s$ and $\mathsf{A}_t$ are one-row matrices, and each $p_v$ is a scalar-valued linear function.
\end{corollary}
\begin{proof}
Apply Theorem~\ref{thm:normal} with reference terminal $s$. Since $(r-1)\ell=1$, write each announcement as a scalar $P_v=\mathsf{B}_vR_{E(v)}$. The decoding identity is
\[
 S=\mathsf{A}_sR_{E(s)}
   =\mathsf{A}_tR_{E(t)}+P_s+P_t+\sum_{v\ne s,t}P_v.
\]
Subtract $P_s$ from the common key. Terminal $s$ can compute this new key using the row $\mathsf{A}_s-\mathsf{B}_s$, and terminal $t$ uses the row $\mathsf{A}_t+\mathsf{B}_t$ together with the relay announcements. Hence neither terminal needs to announce. The new key is a translation by a quantity in the original public transcript and remains uniform and secret even given that transcript. Omitting $P_s,P_t$ cannot reduce secrecy. Renaming the two rows and the key gives \eqref{eq:two_normal}.
\end{proof}

The identity in \eqref{eq:two_normal} relates local linear terms through shared edge variables. Comparing the coefficient of each edge suggests a conservation law, motivating the flow construction in the next section. There we first derive a connectivity condition for arbitrary KRP and then construct a scalar linear SNC directly. The two-terminal equivalence therefore does not require the original KRP to be linear.

\section{Two-Terminal Equivalence via Flow Conservation}
\label{sec:two}
We first consider two key terminals; the graph may contain any number of helpers. In this case, arbitrary key agreement can be realized by scalar linear SNC on a suitable acyclic orientation. We use connectivity as an intermediate criterion. For binary keys, the key-agreement criterion and a flow-based construction have antecedents in the $st$-flow protocol of~\cite{Mizuki2011}. We give the proof for the prescribed finite field $F$ and explain the precise correspondence with that work at the end of the section.

\begin{theorem}[Two terminals and one symbol]
\label{thm:two}
Fix a finite field $F$, an undirected multigraph $G$, terminal set $U=\{s,t\}$ with $s\ne t$, and a wiretap family $\Acal$. Each edge and the key or message have alphabet $F$, and independent private randomness is allowed at every vertex. The following are equivalent:
\begin{enumerate}[label=(\roman*)]
\item An admissible KRP exists.
\item An admissible linear KRP exists.
\item An admissible scalar linear D0 SNC exists with source $s$ and receiver $t$.
\end{enumerate}
No linearity is assumed in (i), and the result is unchanged if $s$ and $t$ are exchanged. In (iii), D0 may be replaced by D1 or D2, and the linearity requirement may be retained or dropped.
\end{theorem}

The implication (iii)$\Rightarrow$(ii) follows from the one-time-pad simulation, and (ii)$\Rightarrow$(i) is immediate. For the converse, we use the following intermediate lemma. Both its necessity and its constructive sufficiency are proved below.

\begin{connectivitylemma}
Under the assumptions of Theorem~\ref{thm:two}, an admissible KRP requires
\begin{equation}
 G-A\text{ contains an $s$--$t$ path for every }A\in\Acal.
 \label{eq:two-connectivity}
\end{equation}
Conversely, \eqref{eq:two-connectivity} suffices for an admissible scalar linear D0 SNC with source $s$ and receiver $t$ over $F$.
\end{connectivitylemma}
Here $G-A$ is the graph obtained by deleting the edges in $A$ while retaining all vertices. All paths in the connectivity condition are undirected.

\subsection{Necessity: A Fully Observed Cut Precludes a Secret Key}
\label{subsec:cutnecessary}
Suppose some $A\in\Acal$ disconnects $s$ and $t$ in $G-A$. Let $B\subseteq V$ be the vertex set of the connected component containing $s$. Its boundary $\delta(B)$ consists of the edges of $G$ with one endpoint in $B$ and the other in $V\setminus B$. Every such edge belongs to $A$, since otherwise its endpoint outside $B$ would be reachable from $s$ in $G-A$.

Put $C=R_{\delta(B)}$. Let $X$ collect the edge variables internal to $B$ and the private randomness of its vertices, and let $Z$ collect the corresponding information internal to $V\setminus B$. The tuples $X$, $Z$, and $C$ partition the independent initial variables. In particular, $X$ and $Z$ are independent conditional on $C$.

This conditional independence persists after conditioning on the entire public transcript $P$. To see this, fix $C=c$ and a transcript $p$ of positive conditional probability. Once all preceding announcements are fixed to their values in $p$, consistency with the next announcement constrains only the initial information of the speaker's side. Multiplying these constraints over the prescribed speaker sequence gives
\[
 \mathbf1_{\{P=p\}}=\alpha(X,c,p)\beta(Z,c,p),
\]
where $\alpha$ and $\beta$ are zero--one functions. Thus the conditional probability of $(X,Z)=(x,z)$, given $(P,C)=(p,c)$, is proportional to
\[
 \bigl(\Pr[X=x\mid C=c]\alpha(x,c,p)\bigr)
 \bigl(\Pr[Z=z\mid C=c]\beta(z,c,p)\bigr).
\]
The two factors depend separately on $x$ and $z$; their normalizing sum also factors. Hence the conditional distribution remains a product, even if the speakers alternate repeatedly between the two sides, and $I(X;Z\mid P,C)=0$.

The two terminals lie on opposite sides of the cut. Perfect agreement therefore makes $S$ a deterministic function of $(X,P,C)$ and also of $(Z,P,C)$. The latter property gives $H(S\mid Z,P,C)=0$, so
\[
 H(S\mid P,C)=I(S;Z\mid P,C)
 \le I(X;Z\mid P,C)=0.
\]
The inequality is conditional data processing, using that $S$ is a function of $(X,P,C)$. Thus the public transcript and the cut variables determine the key. The eavesdropper knows both because $\delta(B)\subseteq A$, whereas perfect secrecy would require
$H(S\mid P,R_A)=H(S)=\log|F|>0$.
This contradiction proves the necessity of \eqref{eq:two-connectivity}, including nonlinear discussion and decoding.

\subsection{Local Encoding by Balancing One Outgoing Edge}
\label{subsec:flowlocal}
Assume \eqref{eq:two-connectivity}. The notation $G-s$ means deletion of the vertex $s$ together with all its incident edges; it is a graph, possibly disconnected. Let $C_t\subseteq V\setminus\{s\}$ be the \emph{vertex set} of its connected component containing $t$, and define the induced subgraph
\[
 H=G[C_t\cup\{s\}].
\]
Every simple $s$--$t$ path, after its initial vertex $s$ is removed, lies in $G-s$ and still leads to $t$. Consequently, all its other vertices belong to $C_t$. Here an internal vertex means a vertex other than the endpoints $s,t$. For every $A\in\Acal$, a simple $s$--$t$ path in $G-A$ is therefore also a path in $H-A$. In addition, $H-s=G[C_t]$ is connected by the definition of $C_t$. Since $\varnothing\in\Acal$, the vertex $s$ is connected to $t$ in $H$, so $H$ itself is connected.

Choose a spanning tree of $H-s$, rooted at $t$. First order the vertices of this tree with every child preceding its parent. The root $t$ is then last. Prepend $s$ to obtain an order of all vertices of $H$. This is a vertex order, not a path: consecutive vertices need not be adjacent, and the tree does not contain $s$. Orient every edge of $H$ from its earlier endpoint to its later endpoint. The resulting directed graph is acyclic because every directed step strictly increases the vertex order.

Every vertex $v\in C_t\setminus\{t\}$ has a tree parent, which comes later, so its edge to that parent is outgoing. The vertex $s$ also has an outgoing edge: it has a neighbor in $H$, and all other vertices follow $s$. Thus every vertex except $t$ has at least one outgoing edge, while $t$ has none. The case $C_t=\{t\}$ is included: the tree consists only of its root, and $s$ is joined to $t$ by at least one edge. Edges outside $H$ are unused.

For each $v\in V(H)\setminus\{t\}$, choose one outgoing edge $e_v$, called its \emph{balancing edge}. On its other outgoing edges, $v$ sends fresh independent uniform symbols, independent also of $M$ and of all randomness at other vertices. It sets the value on the balancing edge to
\begin{equation}
 \begin{split}
 Y_{e_v}={}&\sum_{e\in\In(v)}Y_e
 -\sum_{e\in\Out(v)\setminus\{e_v\}}Y_e\\
 &+\mathbf1_{\{v=s\}}M.
 \end{split}
 \label{eq:flow_encoder}
\end{equation}
Each vertex can perform these operations after receiving all its inputs. In particular, $s$ has no incoming edges and uses only $M$ and its private randomness, as required in D0. For an assignment of edge values, define the divergence at $v$ by
\[
 \divg_vY=\sum_{e\in\Out(v)}Y_e-\sum_{e\in\In(v)}Y_e.
\]
Equation~\eqref{eq:flow_encoder} then gives
\[
 \divg_sY=M,\qquad \divg_vY=0
 \quad(v\in V(H)\setminus\{s,t\}).
\]
In the sum of the divergences over all vertices, every edge value occurs once with each sign, so the sum is zero. It follows that $\divg_tY=-M$. Since $t$ has no outgoing edge, it decodes by
\begin{equation}
 M=\sum_{e\in\In(t)}Y_e.
 \label{eq:flow_decoder}
\end{equation}
This proves perfect decoding. All operations are scalar linear over the prescribed field $F$, and each used edge carries one symbol once.

\begin{figure}[t]
\centering
\begin{tikzpicture}[>=Stealth,thick, every node/.style={font=\small},
 dot/.style={circle,draw=ink,fill=pale,minimum size=8mm}]
\node[dot] (s) at (0,0) {$s$};
\node[dot] (v) at (3,1.1) {$v$};
\node[dot] (w) at (3,-1.1) {$w$};
\node[dot] (t) at (6,0) {$t$};
\draw[->,blue] (s)--node[above,sloped] {$J_s$} (v);
\draw[->,orange] (s)--node[below,sloped] {$M-J_s$} (w);
\draw[->,green] (v)--node[left] {$J_v$} (w);
\draw[->,blue] (v)--node[above,sloped] {$J_s-J_v$} (t);
\draw[->,orange] (w)--node[below,sloped] {$M-J_s+J_v$} (t);
\end{tikzpicture}
\caption{An SNC based on flow conservation, with vertex order $s,v,w,t$. The source generates the private random symbol $J_s$, and the relay $v$ independently generates $J_v$. The two inputs at $t$ sum to $M$. This single code is secure against every wiretap set whose removal leaves an undirected $s$--$t$ path.}
\label{fig:flow}
\end{figure}
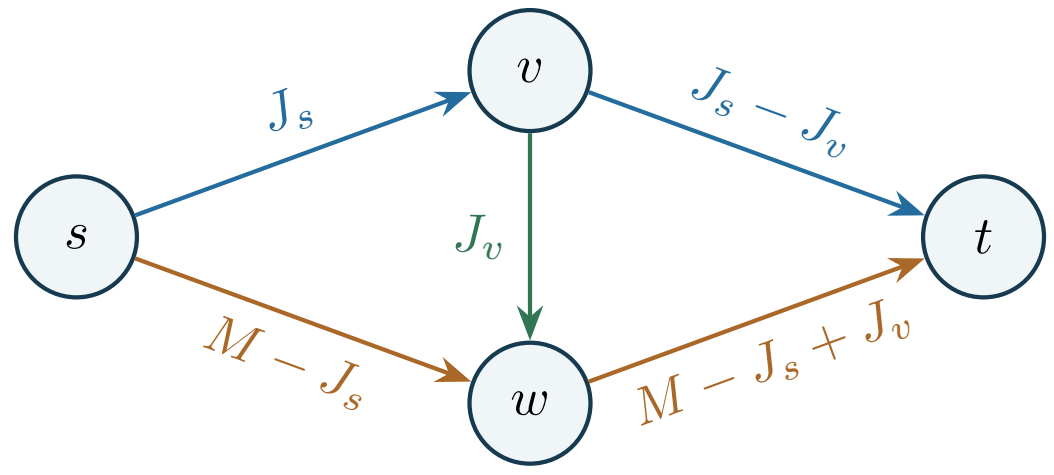

\subsection{Uniformity, Secrecy, and Completion of the Equivalence Proof}
For a fixed message value $M=m$, let
\[
 \mathcal C_m=\left\{y\in F^{E(H)}:
 \begin{array}{l}
 \divg_s y=m,\quad \divg_t y=-m,\\
 \divg_v y=0\quad(v\in V(H)\setminus\{s,t\})
 \end{array}\right\}
\]
be the set of edge assignments satisfying the conservation equations. In the following translations and restrictions, identify each such assignment with its extension by zero on $E\setminus E(H)$.

There is one selected balancing edge $e_v$ for every vertex except $t$. These edges are distinct because a directed edge has only one tail. The other edges carry the freely chosen private random symbols, so their number is
\[
 d=|E(H)|-|V(H)|+1.
\]
This number is nonnegative: the connected graph $H$ contains a spanning tree with $|V(H)|-1$ edges. Once the values on these $d$ edges are chosen, \eqref{eq:flow_encoder} determines the balancing-edge values uniquely in vertex order. The equation at $t$ then follows from the sum of the divergences. Conversely, any $y\in\mathcal C_m$ has those same balancing-edge values when its free coordinates are used as the choices in the encoder. Hence the construction is a bijection from $F^d$ to $\mathcal C_m$. The free coordinates are independent and uniform, so $Y\mid M=m$ is uniform on $\mathcal C_m$. This includes $d=0$, when $F^0$ and $\mathcal C_m$ each have one element.

Now fix $A\in\Acal$ and choose a simple undirected $s$--$t$ path in $H-A$. Define a deterministic vector $z$ indexed by the physical edges: $z_e=1$ if $e$ is on the path and its fixed orientation agrees with traversal from $s$ to $t$, $z_e=-1$ if it disagrees, and $z_e=0$ off the path. The path avoids $A$, so $z_A=0$. At an internal path vertex, one traversal step enters and one leaves. Their signed contributions to the divergence cancel even when the fixed edge orientations do not follow the path. Thus
\begin{gather*}
 z_A=0,\qquad \divg_s z=1,\qquad \divg_t z=-1,\\
 \divg_v z=0\quad(v\in V(H)\setminus\{s,t\}).
\end{gather*}

For any $\delta\in F$, translation by $\delta z$ maps $\mathcal C_m$ to $\mathcal C_{m+\delta}$: it increases the source divergence by $\delta$, decreases the sink divergence by $\delta$, and leaves the other divergences unchanged. Translation by $-\delta z$ is its inverse. Moreover, $(y+\delta z)_A=y_A$ because $z_A=0$. The bijection therefore pairs assignments with exactly the same wiretap observation. Since both conditional distributions are uniform on their respective sets, for every possible observation $a$,
\[
 \Pr[Y_A=a\mid M=m]
 =\Pr[Y_A=a\mid M=m+\delta].
\]
Consequently, the law of $Y_A$ is independent of the message, which is precisely perfect secrecy. It need not be uniform on all of $F^{|A|}$. The argument also applies in characteristic two. The orientation, transmission order, and encoders are fixed for the whole family $\Acal$; only the path used to prove secrecy depends on $A$.

This proves the sufficiency in the connectivity lemma and completes Theorem~\ref{thm:two}. In detail, the lemma gives (i)$\Rightarrow$(iii), Theorem~\ref{thm:otp} gives (iii)$\Rightarrow$(ii), and (ii)$\Rightarrow$(i) is immediate. Every admissible D0, D1, or D2 code, linear or nonlinear, implies (i) by Theorem~\ref{thm:otp}. Conversely, the linear D0 code in (iii) belongs to all three models by \eqref{eq:model-inclusion}. Exchanging $s$ and $t$ in the construction proves the assertion for the other source choice.\hfill$\square$

\subsection{Relation to the Binary st-Flow Results}
Mizuki, Nakayama, and Sone~\cite{Mizuki2011} consider two terminals sharing a uniform bit from independent binary edge keys and public discussion. An exposed edge set $J$ is drawn independently according to a prescribed \emph{leak distribution} $L$. Their security index is the conditional entropy of the shared bit given the eavesdropper's observation, and their objective is to maximize this index~\cite[Def.~6]{Mizuki2011}. Their Lemma~8 proves perfect secrecy for each observation whose exposed edges do not separate the terminals. Their Lemma~10 proves that, for any protocol, an observed cut determines the key. The $st$-flow construction is independent of $L$. These are direct antecedents of the binary key-agreement characterization used here.

To relate the distribution-based formulation to our wiretap family, take $L(A)>0$ for every $A\in\Acal$ and $L(A)=0$ otherwise. For a fixed protocol, independence of $J$ from its initial variables gives
\[
 I(S;P,J,R_J)=\sum_{A\in\Acal}L(A)I(S;P,R_A).
\]
Each term is nonnegative. Thus zero average leakage, equivalently security index one for a uniform bit, holds exactly when the same protocol is perfectly secret for every $A\in\Acal$. The pointwise statements in Lemmas~8 and~10 also yield this conclusion directly. The binary two-terminal cut criterion is therefore not a new claim of this paper.

The construction in~\cite{Mizuki2011} also has an SNC interpretation. Its plaintext edge bits are generated by source randomness, local relay randomness, and XOR conservation on an acyclic orientation; each is sent using its edge key as a one-time pad. Removing these pads and sending the plaintext bits on the physical edges gives a scalar linear D0 code. To check secrecy under this interpretation, write its plaintext vector as $Y$ and its public ciphertext vector as $P=Y+R$, with $R$ independent and uniform. Then $P$ is independent of $(M,Y)$, and $(P,R_A)$ and $(P,Y_A)$ determine one another, so
\[
 I(M;Y_A)=I(M;P,R_A).
\]
Unused edge keys can be ignored. Thus the binary D0 realization can be deduced from the $st$-flow construction, although~\cite{Mizuki2011} does not formulate a comparison between KRP and SNC.

The proof above gives the comparison over any prescribed finite field $F$. It explicitly constructs one acyclic orientation and one scalar linear code that work for the entire wiretap family, with one use of each physical edge and no enlargement of its alphabet. It also supplies the cut converse for nonlinear public discussion. For the threshold family $\Acal=\{A:|A|\le k\}$, \eqref{eq:two-connectivity} says that the minimum $s$--$t$ edge cut has size at least $k+1$; the construction applies equally to nonthreshold families. This two-terminal equivalence provides the reference case for the multi-terminal separations below.

\section{Three Terminals: Separation From D0 and Realization by D1}
\label{sec:triangle}
Let the terminals be $a,b,c$, and join each pair by two parallel edges.
Denote the edge random variables on $ab$, $ac$, and $bc$ by
$X_1,X_2$, $Y_1,Y_2$, and $Z_1,Z_2$, respectively.
The adversary may observe any set of at most two edges.

\begin{theorem}
\label{thm:triangle}
For every finite field $F$, this network admits an admissible linear KRP that shares one symbol of $F$.
For any choice of source, however, it admits no admissible D0 SNC that distributes a message of positive entropy using the same resources.
The impossibility holds even with nonlinear encoding and decoding that uses local randomness.
\end{theorem}
\begin{proof}
The terminals publish
\begin{align*}
 P_a&=X_1-Y_1,& P_b&=X_2-Z_1,& P_c&=Y_2-Z_2.
\end{align*}
Set $T_1=X_1$, $T_2=X_2$, and $T_3=Y_2$. Terminal $a$ already knows all three quantities from its incident edges. Terminal $b$ knows $T_1,T_2$ and obtains $T_3=P_c+Z_2$. Terminal $c$ knows $T_3$ and obtains $T_1=P_a+Y_1$ and $T_2=P_b+Z_1$. Thus every terminal recovers the same key $S=T_1+T_2+T_3$.
The transformation from the six edge random variables to
$(T_1,T_2,T_3,P_a,P_b,P_c)$ is invertible, so these six quantities are also independent and uniform.
Conditioned on the public discussion, the pairs
$\{X_1,Y_1\}$, $\{X_2,Z_1\}$, and $\{Y_2,Z_2\}$ reveal only
$T_1$, $T_2$, and $T_3$, respectively.
Observing two edges reveals at most two of these three independent components.
Their sum $S$ therefore remains perfectly secret.

Suppose that an admissible D0 SNC exists.
Since every vertex is a terminal, the first vertex in a topological order must be the source.
The next vertex has at most two incoming edges, both from the source.
The adversary can observe all of them and thus apply the same decoder as that terminal, contradicting secrecy.
Acyclicity ensures that the received values and $M$ are independent of that terminal's local randomness.
Allowing its decoder to use this randomness does not change the conclusion.
\end{proof}

\begin{figure}[t]
\centering
\begin{tikzpicture}[>=Stealth,thick,every node/.style={font=\small},dot/.style={circle,draw=ink,fill=pale,minimum size=8mm}]
\node[dot] (a) at (0,1.7) {$a=s$};
\node[dot] (b) at (-2,-.6) {$b$};
\node[dot] (c) at (2,-.6) {$c$};
\draw[->,blue] (a) to[bend right=13] node[above left] {$T_1$} (b);
\draw[->,orange] (a) to[bend left=13] node[right,pos=.53] {$T_2$} (b);
\draw[->,blue] (a) to[bend left=13] node[above right] {$T_1$} (c);
\draw[->,green] (a) to[bend right=13] node[left,pos=.53] {$T_3$} (c);
\draw[->,orange] (b) to[bend left=12] node[above] {$T_2$} (c);
\draw[->,green] (c) to[bend left=12] node[below] {$T_3$} (b);
\end{tikzpicture}
\caption{The doubled triangle. The three pairs of equally colored edges form edge-disjoint trees connecting all terminals.
The arrows give a causal D1 SNC: terminal $a$ chooses $T_1,T_2$ independently and uniformly, sets $T_3=M-T_1-T_2$, and sends its four outgoing symbols before the two transmissions between $b$ and $c$.
The directed graph contains a cycle, but the encoding operations have no cyclic dependency.}
\label{fig:triangle}
\end{figure}
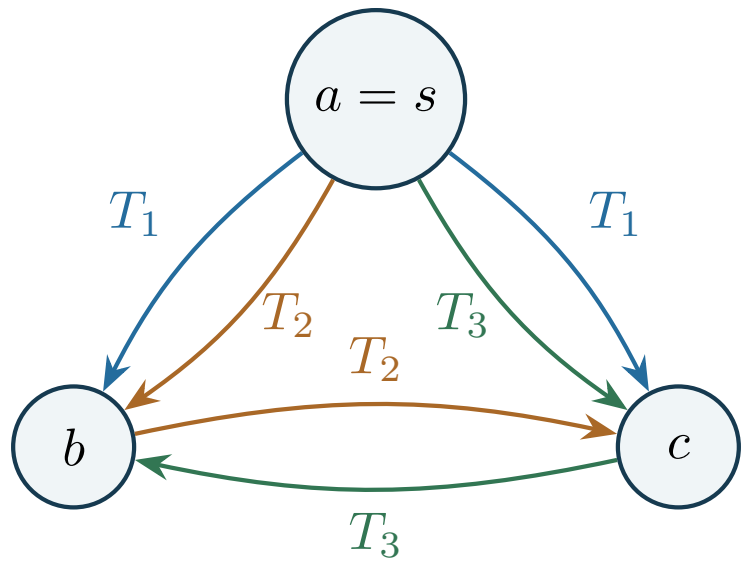

The three equally colored edge pairs in Fig.~\ref{fig:triangle} are edge-disjoint spanning trees. Combining tree-based key sharing is a standard construction for PIN sources~\cite{PIN2010,PerfectPIN2010}. Here its role is to identify exactly which restriction causes the separation.

The causal SNC in Fig.~\ref{fig:triangle} realizes the same task.
All terminals obtain all three $T_i$. For each fixed $M=m$, any two of $T_1,T_2,T_3$ are independent and uniform: the missing component is determined by their sum being $m$, and the two freely chosen components parametrize all $|F|^2$ possibilities bijectively. Thus the distribution of any two components does not depend on $m$. Any two tapped edges reveal at most two components and therefore no information about $M$.
Thus the separation in Theorem~\ref{thm:triangle} arises from the DAG restriction.
It does not extend to encoders that wait only for the inputs they need.
The next section gives a separation that survives this causal extension for linear SNC.

\section{Four Terminals: Separation From Linear Causal SNC}
\label{sec:k4}
The doubled triangle admits a D1 realization and therefore does not separate KRP from causal SNC.
Here we give a linear KRP that cannot be realized by any linear D2 SNC. The impossibility allows independent private randomness at every vertex and its use in decoding, and covers every choice of source, edge orientation, and causal transmission order. The proof identifies the constraints on the coefficient spaces imposed by recovery, secrecy, and causal generation.

\subsection{The Network and Its KRP}
Let all four vertices of the complete graph $K_4$ be terminals, and let $F=\F_3$.
The maximal wiretap sets are the three perfect matchings
\begin{equation}
\begin{aligned}
 A_1&=\{12,34\},& A_2&=\{13,24\},& A_3&=\{14,23\}.
\end{aligned}
\label{eq:k4_sets}
\end{equation}
The family $\Acal$ also contains their subsets. Thus the eavesdropper may observe either edge, or both edges, of one matching. This family does not include every pair of edges.

\begin{theorem}
\label{thm:k4}
This network admits an admissible linear KRP sharing one symbol of $F$.
For every choice of source, however, it admits no admissible linear D2 SNC distributing one symbol of the same field to the other three terminals.
The impossibility includes decoders that use the receiving vertex's local randomness.
\end{theorem}

In the KRP, each terminal publishes one symbol:
\begin{equation}
\begin{aligned}
 P_1&=R_{12}+R_{13}-R_{14},\\
 P_2&=R_{12}+R_{24}-R_{23},\\
 P_3&=R_{34}+R_{13}-R_{23},\\
 P_4&=R_{34}+R_{24}-R_{14}.
\end{aligned}
\label{eq:k4_public}
\end{equation}
Set $X_1=R_{12}$ and $X_2=R_{13}$. The public transcript determines the two scalars
\begin{align*}
 \Delta_1&=(P_1+P_2-P_3-P_4)/2,\\
 \Delta_2&=(P_1-P_2+P_3-P_4)/2,
\end{align*}
where division is in $\F_3$. The edge random variables can then be written as follows.
\begin{center}
\begin{tabular}{ccc}
\toprule
Matching & First edge & Second edge \\
\midrule
$A_1$ & $R_{12}=X_1$ & $R_{34}=X_1-\Delta_1$ \\
$A_2$ & $R_{13}=X_2$ & $R_{24}=X_2-\Delta_2$ \\
$A_3$ & $R_{14}=X_1+X_2-P_1$ & $R_{23}=X_1+X_2-\Delta_2-P_2$ \\
\bottomrule
\end{tabular}
\end{center}
The public values and $X_1,X_2$ are functions of the six edge variables, and the table recovers all six edge variables from them. Hence this change of coordinates is invertible. It follows that
$(P_1,P_2,P_3,P_4,X_1,X_2)$ is uniform on $F^6$, so its six components are mutually independent.
Each terminal is incident to one edge in each of the first two matchings. Using the public corrections in the table, it recovers both $X_1$ and $X_2$, and outputs
\[
 S=X_1-X_2.
\]
Conditional on the public transcript, observing either or both edges of a matching reveals at most one of $X_1$, $X_2$, and $X_1+X_2$.
Over $\F_3$, the coefficient vector $(1,-1)$ of $S$ is linearly independent of each of $(1,0)$, $(0,1)$, and $(1,1)$. Thus $S$ is uniform and independent of each of these observed quantities. Since $X_1,X_2$ are also independent of the public transcript, the KRP satisfies perfect secrecy for every $A\in\Acal$.

\begin{figure}[t]
\centering
\begin{tikzpicture}[thick,every node/.style={font=\small},dot/.style={circle,draw=ink,fill=pale,minimum size=8mm}]
\node[dot] (1) at (0,3.0) {$1$};\node[dot] (2) at (4.4,3.0) {$2$};
\node[dot] (3) at (0,0) {$3$};\node[dot] (4) at (4.4,0) {$4$};
\draw[blue] (1)--node[above] {$X_1$} (2);
\draw[blue] (3)--node[below] {$X_1$} (4);
\draw[orange,dashed] (1)--node[left] {$X_2$} (3);
\draw[orange,dashed] (2)--node[right] {$X_2$} (4);
\draw[green,densely dotted] (1)--node[above,sloped,pos=.23] {$X_1+X_2$} (4);
\draw[green,densely dotted] (2)--node[above,sloped,pos=.23] {$X_1+X_2$} (3);
\end{tikzpicture}
\caption{The three perfect matchings of $K_4$. Each pair with the same color and line style is a maximal wiretap set. The labels give the edge random variables conditioned on public transcript $P=0$. Each terminal can recover $X_1,X_2$, while the eavesdropper observes only one of $X_1$, $X_2$, and $X_1+X_2$.}
\label{fig:k4}
\end{figure}
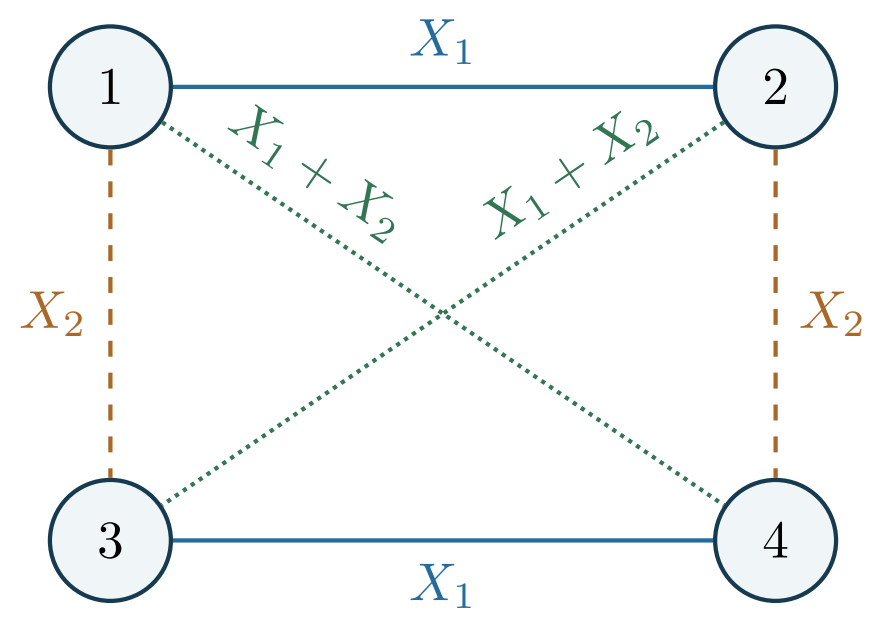

\subsection{Coefficient Spaces Including Local Randomness}
Suppose that an admissible linear D2 SNC exists.
Write the initial information at vertex $v$ as $Z_v=K_v$ for $v\ne s$ and $Z_s=(M,K_s)$. Concatenate these column vectors into $Z\in F^n$, where $n$ is the total number of initial scalar variables.
Every transmitted symbol and the message are linear forms in $Z$, so write
\begin{equation}
 Y_e=\mathbf{g}_eZ,\qquad M=\mathbf{m}Z,
\label{eq:globalcoeff}
\end{equation}
with coefficient row vectors $\mathbf{g}_e,\mathbf{m}\in F^{1\times n}$.
The coordinates of $Z$ are independent and uniform, and $\mathbf{m}\ne0$.
An unused edge has $Y_e=0$ and $\mathbf{g}_e=0$.

We use two kinds of spaces. The spaces $\mathcal Q$ and $\mathcal W_v$ contain coefficient vectors of linear forms in the initial variables. The spaces $\mathcal N$ and $\mathcal N_v$ contain relations among the six edge values:
\begin{equation}
\begin{aligned}
 \mathcal Q&=\Span\{\mathbf{g}_e:e\in E\}\subseteq F^{1\times n},\\
 \mathcal W_v&=\Span\{\mathbf{g}_e:e\in E(v)\}\subseteq F^{1\times n},\\
 \mathcal N&=\left\{\boldsymbol{\lambda}\in F^{1\times6}:\sum_{e\in E}\lambda_e\mathbf{g}_e=0\right\},\\
 \mathcal N_v&=\{\boldsymbol{\lambda}\in\mathcal N:\supp\boldsymbol{\lambda}\subseteq E(v)\}.
\end{aligned}
\label{eq:spaces}
\end{equation}
Here $\lambda_e$ denotes the component of $\boldsymbol{\lambda}$ indexed by edge $e$, and its support is the set of edges with nonzero components.
Membership in $\mathcal N$ means that $\sum_e\lambda_eY_e=0$ for every initial value of $Z$.
Thus $\mathcal W_v$ describes the linear quantities computable from all edges incident to $v$, and $\mathcal N_v$ describes the relations using only those edges.

\begin{lemma}[Two consequences of locality]
\label{lem:local}
For an admissible linear D2 SNC on $K_4$ with every vertex a terminal,
\begin{equation}
 \mathbf{m}\in\mathcal W_v\quad(v\in V),\qquad
 \mathcal N=\sum_{v\in V}\mathcal N_v.
\label{eq:localfacts}
\end{equation}
\end{lemma}
The sum in the second assertion is a sum of subspaces: it consists of all vectors $\sum_v\boldsymbol{\lambda}^{(v)}$ with $\boldsymbol{\lambda}^{(v)}\in\mathcal N_v$. At this stage the sum is not asserted to be direct.
The first assertion says that $M$ can be expressed using all incident edge values, including outgoing values, even when the decoder uses private randomness. The second says that every relation among all edge values is generated by relations supported at individual vertices. Appendix~\ref{app:local} proves both assertions and gives the dimensions and meaning of the local matrices used there.

\subsection{All Transmitted Symbols Lie in a Two-Dimensional Space}
We first determine the dimensions of the spaces above, and then the possible directions of the nonzero edge vectors.

\paragraph{Dimensions at a vertex}
Each vertex has three incident edges, so $\dim\mathcal W_v\le3$.
Lemma~\ref{lem:local} gives $0\ne\mathbf{m}\in\mathcal W_v$. If $\dim\mathcal W_v=1$, any nonzero incident edge vector would be a scalar multiple of $\mathbf{m}$, and that edge value would determine $M$. This contradicts secrecy against a single edge. Thus $\dim\mathcal W_v\ge2$.
The map from the three incident-edge coefficients to their linear combination of $\mathbf{g}_e$ has image $\mathcal W_v$. Extending each input coefficient vector by zero outside $E(v)$ identifies its kernel with $\mathcal N_v\subseteq F^{1\times6}$. Its domain has dimension three, so rank--nullity gives
\begin{equation}
\begin{aligned}
 2&\le\dim\mathcal W_v\le3,\\
 \dim\mathcal N_v&=3-\dim\mathcal W_v\in\{0,1\}.
\end{aligned}
\label{eq:stardims}
\end{equation}
In particular, a vertex contributes a nonzero local relation precisely when its incident-edge coefficient space has dimension two.

\paragraph{At least three vertices have dimension two}
Suppose, to the contrary, that at most two vertices have $\dim\mathcal W_v=2$. Choose distinct vertices $i,j$ containing all such vertices. At every other vertex, \eqref{eq:stardims} gives $\mathcal N_v=\{0\}$, so \eqref{eq:localfacts} reduces to
$\mathcal N=\mathcal N_i+\mathcal N_j$.
Since $\mathbf{m}\in\mathcal W_i\cap\mathcal W_j$, choose row vectors $\mathbf{a}_i,\mathbf{a}_j\in F^{1\times6}$ satisfying
\[
 \supp\mathbf{a}_i\subseteq E(i),\quad
 \supp\mathbf{a}_j\subseteq E(j),\quad
 \sum_e(\mathbf{a}_i)_e\mathbf{g}_e
 =\sum_e(\mathbf{a}_j)_e\mathbf{g}_e=\mathbf{m}.
\]
Subtracting these two representations shows that $\mathbf{a}_i-\mathbf{a}_j\in\mathcal N$.
Hence there exist $\boldsymbol{\lambda}^{(i)}\in\mathcal N_i$ and $\boldsymbol{\lambda}^{(j)}\in\mathcal N_j$ with
$\mathbf{a}_i-\mathbf{a}_j=\boldsymbol{\lambda}^{(i)}+\boldsymbol{\lambda}^{(j)}$.
The corrected coefficient vector
\[
 \mathbf{b}=\mathbf{a}_i-\boldsymbol{\lambda}^{(i)}
          =\mathbf{a}_j+\boldsymbol{\lambda}^{(j)}
\]
is supported on $E(i)$ by its first expression and on $E(j)$ by its second. In $K_4$, the intersection of these sets is the single edge $ij$.
Subtracting or adding a relation does not change the represented linear form, so $\sum_e b_e\mathbf{g}_e=\mathbf{m}$ still holds. Therefore $M=b_{ij}Y_{ij}$, contradicting secrecy against that edge. This proves the assertion.

\paragraph{The coefficient spaces coincide}
Choose three vertices whose spaces $\mathcal W_v$ have dimension two.
If two of these spaces, $\mathcal W_i$ and $\mathcal W_j$, are distinct, both contain the nonzero vector $\mathbf{m}$. Their intersection consequently has dimension one and equals $\Span\{\mathbf{m}\}$.
The common edge vector $\mathbf{g}_{ij}$ belongs to this intersection. If it were nonzero, $Y_{ij}$ would determine $M$. Secrecy therefore forces $\mathbf{g}_{ij}=0$.

If the three spaces were not all equal, one of them would differ from both of the others: choose the unique space if exactly two coincide, or any one if all three are distinct. At its vertex, the two edges joining the other selected vertices would have zero coefficient vectors by the preceding argument. The remaining single incident edge could span a space of dimension at most one, contradicting the choice of a two-dimensional $\mathcal W_v$.
Thus the three spaces coincide.
Every edge of $K_4$ has at least one endpoint among these three vertices, so every $\mathbf{g}_e$ belongs to their common space. This space is therefore $\mathcal Q$, and rank--nullity for the map from six edge coefficients gives
\begin{equation}
 \dim\mathcal Q=2,\qquad \dim\mathcal N=6-2=4.
\label{eq:ranktwo}
\end{equation}
The remaining vertex also has $\mathcal W_v\subseteq\mathcal Q$ and $\dim\mathcal W_v\ge2$ by \eqref{eq:stardims}, so its space equals $\mathcal Q$ as well.
Applying \eqref{eq:stardims} once more, all four spaces $\mathcal N_v$ have dimension one. They sum to $\mathcal N$ by \eqref{eq:localfacts}, and their dimensions sum to $4=\dim\mathcal N$. Hence their sum is direct:
\[
 \mathcal N=\bigoplus_{v\in V}\mathcal N_v.
\]
In particular, distinct local-relation spaces have zero intersection.

\paragraph{Three distinct one-dimensional subspaces}
No edge vector $\mathbf{g}_{ij}$ can be zero. Otherwise, the row vector whose $ij$ component is one and whose other components are zero would be a nonzero relation supported both on $E(i)$ and on $E(j)$. It would belong to $\mathcal N_i\cap\mathcal N_j$, contradicting the direct sum just proved.

Consider the two edges of a wiretap matching. If their coefficient vectors were linearly independent, they would span the two-dimensional space $\mathcal Q$, and the observed values would determine $M$. The two vectors must therefore be linearly dependent. Since neither is zero, they span the same one-dimensional subspace. Denote the subspaces associated with the three matchings by $\mathcal L_1,\mathcal L_2,\mathcal L_3$.

Suppose two of these subspaces coincide, and let $E_0$ be the four edges in their two matchings. The four nonzero coefficient vectors on $E_0$ then have rank one. Each vertex has exactly two incident edges in $E_0$. Its third incident edge has a coefficient vector outside this common one-dimensional subspace, because $\dim\mathcal W_v=2$. Consequently, any local relation must have coefficient zero on that third edge, and every $\mathcal N_v$ is supported on $E_0$.
The relation space of four rank-one vectors has dimension $4-1=3$. It cannot contain the direct sum of the four one-dimensional spaces $\mathcal N_v$, whose dimension is four. This contradiction proves that the three subspaces are distinct. Finally, secrecy against one edge excludes $\mathbf{m}$ from each of them. Thus
\begin{equation}
 \mathcal L_1,\mathcal L_2,\mathcal L_3\text{ are distinct},\qquad
 \mathbf{m}\notin\mathcal L_i\quad(i=1,2,3).
\label{eq:threelines}
\end{equation}
Each vertex is incident to one edge of each of these three types.

\subsection{The Edge Values Cannot Be Generated Causally}
Let the first transmitted symbol be $W=\mathbf{w}Z$.
Every edge vector is nonzero, so every edge is used and $W$ is a nonconstant linear form. Secrecy against its edge makes $M$ and $W$ independent. Their coefficient vectors are consequently linearly independent, and \eqref{eq:ranktwo} gives
$\mathcal Q=\Span\{\mathbf{m},\mathbf{w}\}$.

Before receiving anything, the first sender uses only its initial information. Thus $W$ depends only on the source variables if the source sends first, and only on $K_t$ if a nonsource vertex $t$ sends first.
At any other nonsource vertex, private randomness is independent of $(M,W)$. Such randomness cannot contribute a new linear form in $\mathcal Q$: any quantity in $\mathcal Q$ computable there must lie in the span of the already received quantities. Appendix~\ref{app:local} proves this assertion as an identity between coefficient spaces, and also treats the first sender and the source.

\paragraph{The source sends first}
Consider the first transmission on an edge of the triangle formed by the three nonsource vertices. Such a transmission must eventually occur because every edge is used. Before it occurs, a nonsource vertex can have received only on its single edge to the source. It therefore knows, within $\mathcal Q$, at most the one-dimensional subspace associated with that edge.
At either endpoint of any edge in the nonsource triangle, the triangle edge belongs to a different matching from the source edge. By \eqref{eq:threelines}, its required coefficient vector lies on a different one-dimensional subspace and cannot be computed from the source-edge value. Thus neither endpoint can perform this first triangle transmission, a contradiction.

\paragraph{A nonsource vertex sends first}
Let this vertex be $t$. Then $W$ depends only on $K_t$.
Within $\mathcal Q$, the initial information at $t$ provides precisely $\Span\{\mathbf{w}\}$, the source's initial information provides precisely $\Span\{\mathbf{m}\}$, and the other two vertices provide only zero.

If the first edge does not meet the source, then immediately after it is sent its two endpoints can compute only multiples of $W$ within $\mathcal Q$. At each of those endpoints, the two remaining incident edges belong to different one-dimensional subspaces from the first edge, so neither endpoint can send again. The remaining nonsource vertex has no nonzero quantity in $\mathcal Q$. The source has only multiples of $M$, and \eqref{eq:threelines} excludes this direction from every edge subspace. Thus no vertex can make a second transmission.

If the first edge meets the source, then the source receives $W$ and can compute all of $\mathcal Q$ using $(M,W)$. It may therefore send on its other incident edges. Before the first transmission within the nonsource triangle, however, every nonsource vertex still knows at most its source-edge subspace: vertex $t$ knows the first symbol it sent, and the other vertices can only have received their respective source-edge symbols. The same distinct-subspace argument as in the source-first case makes the first triangle transmission impossible.
Both possibilities lead to a contradiction. This proves the linear D2 impossibility and completes Theorem~\ref{thm:k4}.\hfill$\square$

The obstruction concerns causal generation of the edge values.
The KRP construction yields consistent edge values when its public transcript is fixed, but that consistency does not supply a causal SNC procedure for generating them.
Here recovery and secrecy force all transmitted values into a two-dimensional coefficient space with three distinct one-dimensional edge subspaces.
A nonsource vertex that has learned one type of value cannot generate the different type required for its next edge. Additional private randomness does not enlarge the part of this forced coefficient space that it can compute.

\section{Conclusion}
\label{sec:conclusion}
We studied whether KRP using shared link keys and public discussion, with perfect secrecy against a prescribed family of exposed-edge sets, can be reproduced by SNC on the same undirected physical graph. The SNC designer may choose the source terminal, edge orientations, and causal transmission order. For two terminals and a one-symbol key, arbitrary KRP, linear KRP, and scalar linear D0 SNC are equivalent over the same prescribed finite field $F$. The binary key-sharing characterization and flow construction have antecedents in $st$-flow theory; we made their correspondence with SNC explicit and gave the field-general proof. The construction permits and uses independent randomness at relay nodes.

For more terminals, a single orientation and causal schedule must meet all decoding and secrecy requirements simultaneously. The doubled triangle shows that a DAG restriction can create a separation that disappears when causal cycles are permitted. The four-terminal complete graph shows that allowing causal cycles and receiver randomness does not remove every separation from linear SNC. Its linear KRP shares one ternary symbol, while the linear converse excludes delivery of a source symbol of the same size for every source, orientation, and schedule. Thus, extending the comparison beyond two key terminals and allowing a ternary alphabet exposes a failure of the two-terminal equivalence with linear SNC. The proof identifies a conflict between local linear relations and causal generation of the edge values. Under the stated physical-resource constraints, the standard SNC-to-KRP simulation therefore has no converse that always produces a linear SNC, even for one common key.

The choices of direction and schedule are fixed before execution, and each edge permits at most one transmission. The separation concerns this one-shot task; it does not establish an asymptotic capacity gap or exclude smaller positive multicast rates. Characterizing feasibility for general terminal sets and wiretap families, and determining what remains of the separations under repeated edge use, adaptive schedules, or larger alphabets, are further questions. The normal form proved here applies to linear KRP; elimination of interaction for arbitrary nonlinear KRP remains outside its scope.

\appendices
\section{Proof of the Linear KRP Normal Form}
\label{app:normal}
Write $R_v=R_{E(v)}$ and let $k_v$ be the number of components of the private random vector $K_v$. All random vectors are columns. Sans-serif letters below denote deterministic matrices over $F$.

\paragraph{Removing interaction and grouping announcements}
Let $v_1,\ldots,v_T$ be the speaker sequence, and let $n_t$ be the length of the announcement $P_t$. Linearity gives
\[
 P_t=\mathsf{B}_tR_{v_t}+\mathsf{D}_tK_{v_t}
       +\sum_{j<t}\mathsf{H}_{tj}P_j,
\]
where
\[
 \mathsf{B}_t\in F^{n_t\times|E(v_t)|},\qquad
 \mathsf{D}_t\in F^{n_t\times k_{v_t}},\qquad
 \mathsf{H}_{tj}\in F^{n_t\times n_j}.
\]
Define
\[
 \widetilde P_t=P_t-\sum_{j<t}\mathsf{H}_{tj}P_j
              =\mathsf{B}_tR_{v_t}+\mathsf{D}_tK_{v_t}.
\]
The vector $\widetilde P_t$ can be computed from the speaker's initial information alone. The original transcript determines every $\widetilde P_t$. Conversely, one recovers the original messages successively from
$P_t=\widetilde P_t+\sum_{j<t}\mathsf{H}_{tj}P_j$.
Thus $P\mapsto\widetilde P$ is an invertible block lower triangular transformation, with identity matrices on its diagonal. Replacing one transcript by the other changes neither the information available for decoding nor the eavesdropper's information.

Next stack all blocks $\widetilde P_t$ with speaker $v_t=v$, in their original order, to obtain
\[
 \widehat P_v=(\widetilde P_t)_{t:v_t=v}
             =\widehat{\mathsf{B}}_vR_v
              +\widehat{\mathsf{D}}_vK_v.
\]
The corresponding matrices are obtained by stacking the same rows of $\mathsf{B}_t$ and $\mathsf{D}_t$. Let $p_v=\sum_{t:v_t=v}n_t$ be the length of this grouped announcement. Then
$\widehat{\mathsf{B}}_v\in F^{p_v\times|E(v)|}$ and
$\widehat{\mathsf{D}}_v\in F^{p_v\times k_v}$.
Every vertex can now announce its vector once, in any order. Grouping only permutes the transcript coordinates. For the rest of the proof, suppress the hats and write
\[
 P_v=\mathsf{B}_vR_v+\mathsf{D}_vK_v,
 \qquad P=(P_v)_{v\in V}.
\]
Empty messages and the case $T=0$ are covered by zero-length vectors and the corresponding empty matrices.

\paragraph{Removing private randomness from the decoders}
The decoder at terminal $u$ can be written as
\[
 S=\mathsf{A}_uR_u+\mathsf{E}_uK_u
       +\sum_{v\in V}\mathsf{T}_{uv}P_v,
\]
with
$\mathsf{A}_u\in F^{\ell\times|E(u)|}$,
$\mathsf{E}_u\in F^{\ell\times k_u}$, and
$\mathsf{T}_{uv}\in F^{\ell\times p_v}$.
Choose another terminal $w\ne u$. Its decoder has no separate $K_u$ term, and among the grouped announcements only $P_u$ can depend on $K_u$. Since the two decoders agree for every assignment of the independent initial variables, their coefficients of $K_u$ must agree:
\[
 \mathsf{E}_u+\mathsf{T}_{uu}\mathsf{D}_u
      =\mathsf{T}_{wu}\mathsf{D}_u.
\]
Set $\mathsf{H}_u=\mathsf{T}_{wu}-\mathsf{T}_{uu}\in F^{\ell\times p_u}$. Using $P_u=\mathsf{B}_uR_u+\mathsf{D}_uK_u$, we obtain
\[
 \mathsf{E}_uK_u
      =\mathsf{H}_u\mathsf{D}_uK_u
      =\mathsf{H}_uP_u-\mathsf{H}_u\mathsf{B}_uR_u.
\]
Substitute this identity into the decoder. Define
$\mathsf{A}'_u=\mathsf{A}_u-\mathsf{H}_u\mathsf{B}_u$,
$\mathsf{C}_{uu}=\mathsf{T}_{uu}+\mathsf{H}_u$, and
$\mathsf{C}_{uv}=\mathsf{T}_{uv}$ for $v\ne u$. This gives
\begin{equation}
 S=\mathsf{A}'_uR_u+\sum_{v\in V}\mathsf{C}_{uv}P_v,
 \label{eq:normal_step2}
\end{equation}
where $\mathsf{A}'_u\in F^{\ell\times|E(u)|}$ and
$\mathsf{C}_{uv}\in F^{\ell\times p_v}$. The key has not changed; only its expression has changed, and the decoder no longer needs $K_u$ separately. Apply this argument to every terminal.

\paragraph{Translating the key and retaining the required announcements}
Choose the reference terminal $u_0$ and define
\begin{align*}
 S'&=S-\sum_{v\in V}\mathsf{C}_{u_0v}P_v
       =\mathsf{A}'_{u_0}R_{u_0},\\
 P'_{v,u}&=(\mathsf{C}_{uv}-\mathsf{C}_{u_0v})P_v
       \quad(u\in U\setminus\{u_0\}).
\end{align*}
Subtracting the two versions of \eqref{eq:normal_step2} for $u_0$ and $u$ gives
\[
 \mathsf{A}'_{u_0}R_{u_0}-\mathsf{A}'_uR_u
       =\sum_{v\in V}P'_{v,u}.
\]
Fix a vertex $v$. On the right, only $P'_{v,u}$ can involve $K_v$, whereas the left does not involve any private randomness. Comparing the coefficients of $K_v$ therefore gives
\[
 (\mathsf{C}_{uv}-\mathsf{C}_{u_0v})\mathsf{D}_v=0.
\]
Consequently,
\[
 P'_{v,u}=\mathsf{B}'_{v,u}R_v,\qquad
 \mathsf{B}'_{v,u}
       =(\mathsf{C}_{uv}-\mathsf{C}_{u_0v})\mathsf{B}_v
       \in F^{\ell\times|E(v)|}.
\]
Each vertex can compute these blocks from its incident edge variables alone. Stacking $\mathsf{B}'_{v,u}$ over $u\ne u_0$ gives the matrix $\mathsf{B}'_v$ in Theorem~\ref{thm:normal}. Terminal $u$ decodes using
\[
 S'=\mathsf{A}'_uR_u+\sum_{v\in V}P'_{v,u}.
\]
The key and all new announcements are now functions of the edge variables alone, so the private random vectors can be omitted without changing their joint distribution with those variables.

It remains to check uniformity and secrecy. Fix $A\in\Acal$ and an observation $(P,R_A)=(p,r_A)$ of positive probability. The original key is uniform over $F^\ell$ under this conditioning. The quantity subtracted from $S$ to obtain $S'$ is determined by $p$. Translation by this fixed vector is a permutation of $F^\ell$, so $S'$ is also conditionally uniform. Hence
$I(S';P,R_A)=0$ and $S'$ is uniform over $F^\ell$.
Since the new transcript $P'=(P'_v)_{v\in V}$ is a function of $P$, the data processing inequality gives $I(S';P',R_A)=0$.
Finally, each vertex publishes $r-1$ blocks of length $\ell$, hence at most $(r-1)\ell$ symbols. Writing $\mathsf{A}_u$ for $\mathsf{A}'_u$ gives precisely \eqref{eq:normal}.\hfill$\square$

\section{Local Randomness and Linear Relations Among Edge Values}
\label{app:local}
We prove Lemma~\ref{lem:local} and justify the use of private randomness in the final causality argument.
Let $n_v$ be the number of components of the initial-information vector $Z_v$, so that $\sum_v n_v=n$.
For each used edge, move all previously received symbols in its encoding equation to the left-hand side, leaving the sender's initial variables on the right.
Assign the equation $x_e=0$ for each unused edge to either endpoint.
If $p_v$ equations are assigned to vertex $v$, stacking them gives
\begin{equation}
 \mathsf{L}_v x=\mathsf{A}_v z_v,
\label{eq:local_matrix}
\end{equation}
where $x\in F^6$ is a column vector of edge values, $z_v\in F^{n_v}$ is a possible initial value, and
\[
 \mathsf{L}_v\in F^{p_v\times6},\qquad
 \mathsf{A}_v\in F^{p_v\times n_v}
\]
are deterministic coefficient matrices.
For example, the equation
$x_e=\alpha x_f+\beta x_g+\gamma z_{v,1}$ contributes a row of $\mathsf{L}_v$ with entries $1,-\alpha,-\beta$ in coordinates $e,f,g$, respectively, and zeros elsewhere. The corresponding row of $\mathsf{A}_v$ has first entry $\gamma$ and all other entries zero. Here $e$ is sent by $v$, and $f,g$ are edges received earlier at $v$, so every row of $\mathsf{L}_v$ is supported on $E(v)$.
All actual random vectors satisfy $\mathsf{L}_vY_E=\mathsf{A}_vZ_v$.

\subsection{Every Relation Is Generated by Local Relations}
Let $\mathcal Y\subseteq F^6$ be the space of edge-value vectors attainable by varying all initial values.
An attainable $x$ satisfies $\mathsf{L}_vx\in\im\mathsf{A}_v$ for every $v$ by \eqref{eq:local_matrix}.
Conversely, if these inclusions hold, choose $z_v$ with $\mathsf{A}_vz_v=\mathsf{L}_vx$ at each vertex. These choices can be made independently because the initial-variable blocks have unrestricted Cartesian-product support. In the first transmission, the encoding equation then produces the prescribed component of $x$. If all earlier transmissions have their prescribed values, the next encoding equation does so as well. Induction along the fixed transmission order proves that the entire vector $x$ is attained.

For each $v$, choose a matrix $\mathsf{B}_v$ whose rows form a basis of the left nullspace of $\mathsf{A}_v$. Its dimensions are
\[
 \mathsf{B}_v\in F^{q_v\times p_v},\qquad
 q_v=p_v-\operatorname{rank}\mathsf{A}_v,
\]
and $\ker\mathsf{B}_v=\im\mathsf{A}_v$.
The preceding characterization therefore gives
\[
 \mathcal Y=\{x\in F^6:\mathsf{B}_v\mathsf{L}_vx=0\text{ for every }v\}.
\]
A row vector that vanishes on this common kernel belongs to the row space of the vertically stacked matrices $\mathsf{B}_v\mathsf{L}_v$. Thus every relation in $\mathcal N$ is a linear combination of their rows. Each row is supported on $E(v)$ and vanishes on every attainable edge vector, so it belongs to $\mathcal N_v$.
This proves $\mathcal N\subseteq\sum_v\mathcal N_v$. The reverse inclusion follows from the definition $\mathcal N_v\subseteq\mathcal N$, establishing the second assertion of Lemma~\ref{lem:local}.

\subsection{The Message Is Determined by All Incident Edges}
First vary the source's initial value $z_s$ by any vector in $\ker\mathsf{A}_s$, while fixing every other initial value. All right-hand sides in \eqref{eq:local_matrix} remain unchanged. Causality, by the same induction as above, implies that all edge values remain unchanged. Every other terminal has unchanged local randomness and received values, so its decoded message also remains unchanged. Perfect decoding for all initial values consequently forces the source message to remain unchanged.

Let $\boldsymbol{\mu}_s\in F^{1\times n_s}$ be the row vector extracting the message coordinate from $Z_s$, so $M=\boldsymbol{\mu}_sZ_s$.
The preceding variation shows that $\boldsymbol{\mu}_s$ vanishes on $\ker\mathsf{A}_s$. It therefore belongs to the row space of $\mathsf{A}_s$: for some $\mathbf{b}_s\in F^{1\times p_s}$,
\[
 \boldsymbol{\mu}_s=\mathbf{b}_s\mathsf{A}_s,
 \qquad M=\mathbf{b}_s\mathsf{A}_sZ_s
          =\mathbf{b}_s\mathsf{L}_sY_E.
\]
Since every row of $\mathsf{L}_s$ is supported on $E(s)$, this expresses $M$ using only the incident edges at the source.

For a nonsource vertex $v$, write its linear D2 decoding equation as
\[
 M=\mathbf{c}_vY_E+\mathbf{d}_vZ_v,
 \qquad \mathbf{c}_v\in F^{1\times6},\quad
 \mathbf{d}_v\in F^{1\times n_v},
\]
where $\mathbf{c}_v$ is supported only on incoming edges.
Varying $z_v$ in $\ker\mathsf{A}_v$ leaves all edge values unchanged by causality and leaves $M$ unchanged because $v$ is not the source. The decoding identity then implies that $\mathbf{d}_v$ vanishes on $\ker\mathsf{A}_v$. Hence $\mathbf{d}_v=\mathbf{b}_v\mathsf{A}_v$ for some row vector $\mathbf{b}_v\in F^{1\times p_v}$, giving
\[
 M=(\mathbf{c}_v+\mathbf{b}_v\mathsf{L}_v)Y_E.
\]
Both terms in the coefficient row are supported on $E(v)$. Thus at every vertex $M$ is a linear combination of its incident edge values, which is exactly $\mathbf{m}\in\mathcal W_v$.
This recovers the contribution of private randomness required for decoding, not necessarily the entire private-randomness vector.\hfill$\square$

\subsection{Local Randomness in the Causality Argument}
Let $\mathcal I_v\subseteq F^{1\times n}$ be the span of the coordinate row vectors belonging to the initial block $Z_v$. Let $\mathcal R_v\subseteq\mathcal Q$ be the span of the coefficient vectors of the symbols received at $v$ up to the time in question.
The coefficient vectors of linear quantities computable at $v$ then form $\mathcal I_v+\mathcal R_v$. In particular, the computable quantities that belong to $\mathcal Q$ have coefficient space
\[
 (\mathcal I_v+\mathcal R_v)\cap\mathcal Q
   =(\mathcal I_v\cap\mathcal Q)+\mathcal R_v.
\]
Indeed, if $\mathbf{i}+\mathbf{r}\in\mathcal Q$ with $\mathbf{i}\in\mathcal I_v$ and $\mathbf{r}\in\mathcal R_v$, then $\mathbf{r}\in\mathcal Q$ implies $\mathbf{i}\in\mathcal Q$. The reverse inclusion is immediate.
Previously sent values add no new initial information: each was already a function of the same initial block and previously received values.

In Section~\ref{sec:k4}, $\mathcal Q=\Span\{\mathbf{m},\mathbf{w}\}$, where $W=\mathbf{w}Z$ is the first transmitted symbol. If the source sends first, both $\mathbf{m}$ and $\mathbf{w}$ are supported on its initial block. Disjointness of the initial-variable blocks gives $\mathcal I_v\cap\mathcal Q=\{0\}$ at every nonsource vertex.
If a nonsource vertex $t$ sends first, then $\mathbf{w}$ is supported on the block of $t$ and $\mathbf{m}$ on that of the source. Both vectors are nonzero. Therefore
\[
 \mathcal I_t\cap\mathcal Q=\Span\{\mathbf{w}\},\qquad
 \mathcal I_s\cap\mathcal Q=\Span\{\mathbf{m}\},
\]
while the corresponding intersections are zero at the other vertices.
These identities account for all private randomness in the causality argument, including random coordinates that never appear in a transmitted symbol.

\section*{Acknowledgment}
This work originated from a research project conducted during the Graduate-level Research in Industrial Projects for Students (G-RIPS Sendai 2025) program. The authors thank Maho Maruyama for her contributions to the project during the program.

ChatGPT (OpenAI) assisted with the development and checking of mathematical arguments, manuscript organization, English drafting and Japanese translation, and preparation of LaTeX and TikZ source throughout Sections~I--VIII and Appendices~A--B. Responsibility for the content rests with the authors.

\end{document}